\documentclass[12pt]{article}
\usepackage{amsmath,amssymb,graphicx,slashbox}
\usepackage{color}

\def\mydate{February 24, 2010}
\def\ignore#1{{}}

\ignore{
\textheight=24.5cm
\textwidth=16cm
\topmargin=-1.5cm
\oddsidemargin=0.0cm
\evensidemargin=0.0cm
}

\newcommand{\bea}{\begin{eqnarray}}
\newcommand{\eea}{\end{eqnarray}}

\makeatletter
\@addtoreset{equation}{section}
\makeatother

\newcommand{\beeq}{\begin{equation}}
\newcommand{\eneq}{\end{equation}}
\newcommand{\beqn}{\begin{eqnarray}}
\newcommand{\eeqn}{\end{eqnarray}}

\def\dd{\partial}
\def\la{\raise.16ex\hbox{$\langle$}\lower.16ex\hbox{}  }
\def\ra{\raise.16ex\hbox{$\rangle$}\lower.16ex\hbox{} }

\def\onehalf{ \hbox{${1\over 2}$} }

\def\eff{{\rm eff}}

\def\EM{{\rm EM}}

\def\KK{{\rm KK}}

\def\psibar{ \psi \kern-.65em\raise.6em\hbox{$-$} }
\def\psibarl{ \psi \kern-.65em\raise.6em\hbox{$-$} \lower.6em\hbox{} }

\def\myred#1{{#1}}
\def\myblue#1{{#1}}

\begin{document}

\thispagestyle{empty}

{\small \noindent \mydate    \hfill OU-HET 649/2009}

\vspace{4.0cm}

\baselineskip=35pt plus 1pt minus 1pt

\begin{center}
{\LARGE \bf 
The electroweak gauge couplings \\
in SO(5)$\times$U(1) gauge-Higgs unification
}
\end{center}

\vspace{2.0cm}
\baselineskip=20pt plus 1pt minus 1pt

\begin{center}
{\bf  
Yutaka Hosotani, Shusaku Noda and Nobuhiro Uekusa
}


{\small \it Department of Physics, 
Osaka University, 
Toyonaka, Osaka 560-0043
Japan} \\
\end{center}


\vskip 2.5cm
\baselineskip=20pt plus 1pt minus 1pt

\begin{abstract}

The electroweak  currents of quarks and leptons 
in the $SO(5) \times U(1)$ gauge-Higgs unification model 
in the Randall-Sundrum warped space are determined.
The 4D gauge couplings deviate from those in the standard model
and the weak universality is slightly violated.
It is shown that the model is \myred{free from 4D anomalies} and 
the deviations of the gauge couplings are tiny, 
less than 1\%  except for the top quark.  The $Zt_L \bar t_L$, 
$Zt_R \bar t_R$,  $Zb_L \bar b_L$ and $Zb_R \bar b_R$ couplings
deviate from those in the standard model by  \myred{$-7$\%}, 18\%, 0.3\% and 0.9\%
\myred{with the warp factor $z_L=10^{15}$}, respectively.
The violation of the $\mu$-$e$, $\tau$-$e$ and $t$-$e$ universality in the charged current 
interactions is  ${\cal O}(10^{-8})$, ${\cal O}(10^{-6})$ and 2.3\%, respectively.
\end{abstract}



\newpage

\baselineskip=20pt plus 1pt minus 1pt

\section{Introduction}
The Higgs boson is the only particle yet  to be found 
in the standard model of electroweak interactions.  
It is not clear, however,  if the Higgs boson appears as described in the standard model.
New physics may be hidden behind it.

In the gauge-Higgs unification scenario the 4D Higgs field is identified with 
a part of the extra-dimensional component of gauge fields in higher 
dimensions.\cite{Fairlie1, Manton1}
The 4D Higgs field appears as an Aharonov-Bohm (AB) phase, or a Wilson line phase,
in the extra dimension.\cite{YH1, YH2, Davies1}
The electroweak (EW) symmetry breaking is induced
by dynamics of the AB phase through the Hosotani mechanism.
A finite mass of the Higgs boson is generated at the quantum level, 
the mechanism of which provides a new way of solving the gauge hierarchy problem
as an alternative to supersymmetric theories, the little Higgs model and 
the Higgsless model.\cite{Lim1} 

A realistic gauge-Higgs unification model is constructed in the Randall-Sundrum
warped space.\cite{Pomarol2,  Oda1, HM}
Based on the gauge group $SO(5) \times U(1)$ quarks and leptons
are introduced in the vector ({\bf 5}) representation of $SO(5)$ in the bulk
five-dimensional spacetime with additional fermions localized on the Planck 
brane.\cite{Agashe1}-\cite{HK}
The presence of the top quark dynamically
induces the EW symmetry breaking, thereby the effective potential $V_\eff$ for 
the AB phase $\theta_H$ being minimized at 
$\theta_H = \pm \onehalf \pi$.\cite{HOOS}

In the gauge-Higgs unification scenario the interactions of the Higgs boson are
governed by the gauge principle.  Its interactions  
with other particles deviate from those in the standard model,
which may be understood as manifestation of the underlying gauge invariance.
The 4D neutral Higgs field $H(x)$ corresponds to four-dimensional fluctuations of
the AB phase $\theta_H$ so that these two appear, in the effective theory at low energies,
always in the combination of 
\beeq
\hat \theta_H ( x)  = \theta_H + \frac{H(x)}{f_H}
\label{AB1}
\eneq
where $f_H$ turns out $\sim 246 \,$GeV.
The effective  interactions with   $W$, $Z$ bosons and fermions are 
summarized as \cite{HK}
\beeq
{\cal L}_\eff  = - V_\eff (\hat \theta_H) 
                 - m_W^2(\hat \theta_H) W^\dagger_\mu W^\mu
                 - \onehalf m_Z^2(\hat \theta_H) Z_\mu Z^\mu 
                 - \sum_{a,b} m^F_{ab}(\hat \theta_H) \psibar_a  \psi_b ~.
\label{effective1}
\eneq
The mass functions are approximately given by
$m_W(\hat\theta_H) \sim \onehalf g f_H \sin \hat\theta_H$,
$m^F_f (\hat\theta_H) \sim y_f f_H \sin \hat\theta_H$ etc.
In the standard model  they are given 
by $m_W = \onehalf g (v+H)$ and $m^F_f = y_f (v+H)$.
The periodicity in $\hat \theta_H$ in the gauge-Higgs unification 
follows from the gauge invariance of the theory.

\myblue{
The large gauge invariance implying the periodicity in $\hat \theta_H$,}
\myred{ and the smooth 
$\hat \theta_H$-dependence of the mass functions without the level crossing
in the warped space}  lead to significant deviation in the Higgs
couplings from the standard model.\cite{HK}-\cite{Grojean2}
As follows from (\ref{effective1}), 
the $WWH$, $ZZH$ and Yukawa couplings are suppressed by a factor 
$\sim \cos \theta_H$ compared with those in the standard model.  
At particular values of $\theta_H = \pm \onehalf \pi$, 
\myred{namely at the extrema of the  smooth periodic mass functions,}  
they  vanish.\footnote{\myred{In models in flat spacetime the mass functions can be linear
in $\hat\theta_H$, which results in the level crossing in conformity with the periodicity
in $\hat\theta_H$.  In such a case $\hat\theta_H = \pm \onehalf \pi$ is not realized as 
demonstrated in ref. \cite{HOOS}.}}
As proven in ref.\ \cite{HKT}   the $WWH^\ell$, $ZZH^\ell$ and
$\psibar \psi H^\ell$  couplings with an odd integer $\ell$ 
vanish to all order in perturbation theory, provided that
all bulk fermions belong to the vector representation of $SO(5)$.  
The Higgs boson becomes absolutely stable, the stability being
protected by the  dynamically emerging  $H$-parity.
The Higgs boson is $H$-parity odd, while all other particles in the
standard model  are $H$-parity even.
This leads to astonishing physical consequences. 
In the evolution of the universe, Higgs bosons become the cold dark matter.
The dark matter density  observed at WMAP is explained with $m_H \sim 70  \,$GeV.  
This scenario of the stable Higgs bosons as cold dark matter is quite different 
from the Kaluza-Klein (KK) dark matter scenario in which additional fields 
with odd KK parity become dark matter.\cite{KKdark1}
The LEP2 bound for $m_H$ is evaded as the $ZZH$ coupling vanishes.
Higgs bosons can be produced in pairs in collider experiments .  
They appear as missing energies and momenta.

In this paper we turn our attention to the gauge couplings of quarks and 
leptons.\cite{Agashe3, Wagner2}
Although the five-dimensional gauge couplings are dictated by the gauge principle 
and are universal, the four-dimensional gauge couplings appear as overlap 
integrals in the fifth coordinate with wave functions of relevant fields inserted.
In the gauge-Higgs unification models in the RS warped space each of the left- and 
right-handed quarks and leptons has a quite different profile of the wave function so that
deviation from the standard model is expected to arise in gauge couplings as well.
It has been argued that in the $SU(3)$  model 
the weak ($\mu$-$e$, $\tau$-$e$) universality is slightly violated.\cite{HNSS}
There appear  corrections to the $S$ and $T$ parameters,\cite{Agashe1, Carena1, Lim3}
which would constrain the models. 
Consequences in the tree level unitarity in the $WW$, $WZ$, and $ZZ$ scattering also 
have been discussed.
\cite{Csaki:2003dt}-\cite{unitarity}
Further, it has been shown that corrections to muon anomalous magnetic 
moment   and  to neutron electric dipole moment   
appear in the $SU(3)$ model, 
which has been used to get a bound for the KK mass scale.\cite{Lim4, ALM}

The purpose of this paper is to determine the $W$, $Z$ and electromagnetic 
currents in the $SO(5) \times U(1)$ model with three generations of quarks and
leptons,  which generalizes the model of  ref.\ \cite{HOOS}.    
The electroweak currents  depend on the profiles of wave functions of $W$, $Z$, 
quarks and leptons both in the fifth dimension and in the $SO(5)$ group.
Despite their highly nontrivial profiles, it is found that 
the deviations of the couplings of quarks and leptons to the gauge bosons
from the standard model are less than 1\% except for the top quark.

The paper is organized as follows.  In Section 2 the $SO(5) \times U(1)$ model is specified.
Quark and lepton multiplets are introduced in the bulk five-dimensional spacetime,
with associated  fermions localized on the Planck brane.  
In Section 3 we show that with both quark and lepton multiplets included the
model becomes  \myred{free from 4D anomalies}  with respect to the
 $SO(4) \times U(1)$ gauge symmetry 
left after the orbifold conditions imposed.  Brane fermions play an important role there.
In Section 4 the spectrum of gauge bosons and fermions is determined, and
the wave functions of the photon, $W$, $Z$, quarks, and leptons are 
determined.  Numerical values of the coefficients in the wave functions are
give at $\theta_H = \onehalf \pi$.  
By making use of the wave functions,  the electroweak gauge couplings are
evaluated in Section 5.  Comparison with those in the standard model is given.
Section 6 is devoted to discussions and conclusion.  Useful formulas
are collected in  appendices A, B, and C.   The results for the gauge couplings
evaluated with the pole masses of quarks and leptons are given in the main text, while 
the results with the running quark and lepton masses at the $m_Z$ scale are 
summarized in Appendix D.

\section{Model}

The model is defined in the Randall-Sundrum  (RS) warped spacetime
whose metric is given by
\bea
    ds^2 = G_{MN}dx^M dx^N
  = e^{-2\sigma(y)}\eta_{\mu\nu} dx^\mu dx^\nu + dy^2 , \qquad
      \label{metric1}
\eea
where $\eta_{\mu\nu} =\textrm{diag}(-1,1,1,1)$,
$\sigma(y)=\sigma(y+2L)$, and
$\sigma(y)=k|y|$ for $|y|\leq L$.
The fundamental region in the fifth dimension
is given by $0\leq y\leq L$.
The Planck brane and the TeV brane are located at $y=0$ 
and $y=L$, respectively.  
The bulk region $0 < y < L$ is  an anti-de Sitter spacetime with the 
cosmological constant  $\Lambda = - 6k^2$.


We consider an $SO(5) \times U(1)_X$ gauge
theory in the RS warped spacetime. 
The $SO(5) \times U(1)_X$  symmetry is broken  to $SO(4) \times U(1)_X$
by the orbifold boundary conditions at the Planck and TeV branes.
The symmetry is spontaneously broken to $SU(2)_L \times U(1)_Y$ by additional
interactions at the Planck brane. 
\ignore{
Here we address neither a question of how the orbifold structure 
of spacetime appears with orbifold conditions,
nor a question of how the symmetry further reduces to
the standard model symmetry $SU(2)_L\times U(1)_Y$
on the Planck brane.
We imagine these happen at a high energy 
scale of ${\cal O}(M_{\textrm{\scriptsize GUT}})$
to ${\cal O}(M_{\textrm{\scriptsize Planck}})$.
}

The action integral consists of four parts:
\bea
  S =
  S_{\textrm{\scriptsize bulk}}^{%
  \textrm{\scriptsize gauge}}
  +S_{\textrm{\scriptsize Pl. brane}}^{%
  \textrm{\scriptsize scalar}}
  +S_{\textrm{\scriptsize bulk}}^{%
  \textrm{\scriptsize fermion}}
  +S_{\textrm{\scriptsize Pl. brane}}^{%
  \textrm{\scriptsize fermion}} .
\eea
The bulk parts respect $SO(5)\times U(1)_X$ gauge
symmetry. There are 
$SO(5)$ gauge fields $A_M$ and $U(1)_X$
gauge field $B_M$. The former are decomposed as 
$A_M = \sum_{I=1}^{10} A_M^I T^I 
    = \sum_{a_L=1}^3 A_M^{a_L} T^{a_L}
     +\sum_{a_R=1}^3 A_M^{a_R} T^{a_R}
     +\sum_{\hat{a}=1}^4 A_M^{\hat{a}} T^{\hat{a}}$,
where  
$T^{a_L,a_R}(a_L,a_R=1,2,3)$ and
$T^{\hat{a}}(\hat{a}=1,\ldots,4)$ 
are the generators of
$SO(4)\sim SU(2)_L \times SU(2)_R$ and $SO(5)/SO(4)$,
respectively.
In a vectorial representation, the components of the 
generator are
$T_{ij}^{a_L,a_R}
   =
     -{i\over 2}
      [{1\over 2} \epsilon^{abc}
       (\delta_i^b \delta_j^c
       -\delta_j^b \delta_i^c
       )
     \pm 
     (\delta_i^a \delta_j^4
     -\delta_j^a \delta_i^4)
    ]$
and 
$T_{ij}^{\hat{a}}
  =-{i\over \sqrt{2}}
   (\delta_i^{\hat{a}}
   \delta_j^5
   -\delta_j^{\hat{a}}
    \delta_i^5)$,
where $i,j=1,\ldots, 5$.  They satisfy $\textrm{Tr}(T^I T^J)=\delta^{IJ}$.
The action integral for the pure gauge boson part is 
\bea
   S_{\rm bulk}^{\rm gauge} 
  &\!\!\!=\!\!\!& \int d^5x \sqrt{-G}
  \left[ -\textrm{tr}( {1\over 4} F^{(A)MN} F_{MN}^{(A)}
            +{1\over 2 \xi}
  (f_{\textrm{\scriptsize gf}}^{(A)})^2 + 
  {\cal L}_{\textrm{\scriptsize gh}}^{(A)})
  \right.
\nonumber
\\
  &&\qquad \left.
   -({1\over 4} F^{(B)MN}F_{MN}^{(B)} 
  +{1\over 2\xi}
 (f_{\textrm{\scriptsize gf}}^{(B)})^2
    +{\cal L}_{\textrm{\scriptsize gh}}^{(B)})
   \right] ,
\eea
where the gauge fixing and ghost terms are
denoted as functionals with suffices gf and gh, respectively. Here 
$F_{MN}^{(A)} =
 \partial_M A_N -\partial_N A_M -ig_A  [A_M, A_N]$ 
and 
$F_{MN}^{(B)} =\partial_M B_N -\partial_N B_M$.

The orbifold boundary conditions
at $y_0=0$ and $y_1=L$ for gauge fields are given by
\bea
  && \left(\begin{array}{c}
    A_\mu \\
    A_y \\
   \end{array}\right) (x,y_j-y)
  = P_j \left(\begin{array}{c}
    A_\mu \\
    -A_y \\
   \end{array} \right) (x,y_j+y) P_j^{-1} ,
\nonumber
\\
   && \left(\begin{array}{c}
    B_\mu \\
    B_y \\
   \end{array} \right) (x,y_j-y)
   =
    \left(\begin{array}{c}
     B_\mu \\
     -B_y \\
   \end{array}\right) (x,y_j+y) ,
\nonumber
\\
 && P_j =\textrm{diag}(-1,-1,-1,-1,+1) ,
 \qquad (j=0,1),
 \label{pj}
\eea 
which reduce the $SO(5)\times U(1)_X$ symmetry to
$SO(4)\times U(1)_X$.
A scalar field $\Phi(x)$ on the Planck brane
belongs to $(0,{1\over 2})$ representation
of $SO(4)\sim SU(2)_L \times SU(2)_R$  with a $U(1)_X$ charge.
With the brane action
\bea
  S_{\textrm{\scriptsize Pl. brane}}^{%
   \textrm{\scriptsize scalar}}
 &\!\!\!=\!\!\!&\int d^5x \delta(y)
  \left\{
     -(D_\mu \Phi)^\dag D^\mu \Phi
     -\lambda_\Phi
     (\Phi^\dag \Phi -w^2)^2 \right\} ,
\nonumber
\\
  D_\mu \Phi &\!\!\!=\!\!\!&
   \partial_\mu \Phi 
   -i\left(g_A \sum_{a_R}^3
     A_\mu^{a_R} T^{a_R}
      +{g_B\over 2}B_\mu \right) \Phi ,
      \label{cov}
\eea
the $SU(2)_R\times U(1)_X$ symmetry breaks down to $U(1)_Y$,  
the weak hypercharge in the standard model.
Let us denote
\beqn
&&\hskip -1cm 
   \left(\begin{array}{c}
    A_M^{'3_R} \\
    A_M^Y \\
   \end{array}\right) 
   = \left(\begin{array}{cc}
    c_\phi & -s_\phi \\
    s_\phi & c_\phi \\
   \end{array}\right) 
  \left(\begin{array}{c}
    A_M^{3_R} \\
   B_M \\
  \end{array} \right) \cr
  \noalign{\kern 10pt}
&&\hskip -1cm
    c_\phi ={g_A\over \sqrt{g_A^2 + g_B^2}} ,
  \quad
    s_\phi ={g_B\over \sqrt{g_A^2 + g_B^2}}  ~.
 \label{hyperY1}
\eeqn
Suppose that $w$ is much larger than the  KK mass scale $M_\KK$, being of, for instance,  
${\cal O}(M_{\textrm{\scriptsize GUT}})$ to ${\cal O}(M_{\textrm{\scriptsize Planck}})$.
\myred{Although the fields $A_\mu^{1_R}$, $A_\mu^{2_R}$ and $A_\mu^{'3_R}$
are even under parity  and obey the Neumann condition, the values at $y=0$ of low-lying 
modes of their KK towers  become extremely tiny because of the large VEV $w$. }
The net effect for the low-lying modes of the KK towers of $A_\mu^{1_R}$, $A_\mu^{2_R}$
and $A_\mu^{'3_R}$ is  that they effectively
obey Dirichlet boundary conditions at the Planck 
brane so that the originally-massless modes of $A_\mu^{1_R}$, $A_\mu^{2_R}$ and
$A_\mu^{'3_R}$ acquire large masses of ${\cal O}(M_\KK)$.\cite{HS2}
\myred{On the other hand the fields $A_y^{1_R}$, $A_y^{2_R}$ and $A_y^{'3_R}$
are odd under parity and necessarily vanish at $y=0$.}
The effective orbifold boundary conditions are tabulated in Table~\ref{tab:bc}.
\myred{
In passing, it is possible to allow discontinuities in 
$A_y$ at $y=0$ by enlarging the configuration space for $A_y$ as discussed
in ref.\ \cite{TASI1}.    One can include discontinuities in $A_y$ in the 
gauge-fixing condition at the Planck brane.  Then $A_y$ can obey either Dirichlet,  
or Neumann, or other boundary condition, depending on the gauge condition,
as  spelled out in refs.\ \cite{HOOS} and \cite{TASI1}.
We adopt the viewpoint that all vector potentials $A_M$ are continuous,
but the results in the present paper are not affected by the gauge choice.}

\begin{table}[b]
\begin{center}
\caption{Boundary conditions of gauge fields and bulk fermions: 
The effective Dirichlet condition resulting from brane dynamics
is denoted as D$_{\eff}$.   
For fermions  $\psi_{aj}$, $j=1,\cdots,4$.
\label{tab:bc}} 
\begin{tabular}{|cccccc|c|}
\noalign{\kern 10pt}
\hline 
$A_\mu^{a_L}$ & $ A_\mu^{1_R,2_R}$ & $A_\mu^{'3_R}$ & $A_\mu^Y$ 
& $A_\mu^{\hat{a}}$ & $B_\mu$ 
& $\psi_{ajL},\psi_{a5R}$ \\ \hline
(N,N) & (D$_{\eff}$,N) & 
(D$_{\eff}$,N) & (N,N) 
& (D,D) & (N,N)  & (N,N) \\ \hline 
\noalign{\kern 4pt}
\hline
$A_y^{a_L}$ & $ A_y^{1_R,2_R}$ & $A_y^{'3_R}$ & $A_y^Y$ & $A_y^{\hat{a}}$ & $B_y$ 
&$\psi_{ajR},\psi_{a5L}$ \\ \hline
(D,D) & (D,D) & (D,D) & (D,D) & (N,N) & (D,D)
& (D,D) \\ \hline
\end{tabular}
\end{center}
\end{table}

Bulk fermions for quarks and leptons are introduced as multiplets in 
the vectorial representation of $SO(5)$.
In the quark sector two vector multiplets are introduced for each 
generation.  In the lepton sector it suffices to introduce one multiplet
for each generation to describe massless neutrinos, whereas it is
necessary to introduce two multiplets to describe massive neutrinos.
They are denoted by $\Psi_a^t =(\psi_{a1},\ldots,\psi_{a5})^t$
where the subscript $a$ runs from 1 to 3 or 4 for each generation.

The action integral in the bulk is given by
\bea
   S_{\textrm{\scriptsize bulk}}^{%
     \textrm{\scriptsize fermion}}
   &\!\!\!=\!\!\!& \int d^5 x \,  \sqrt{-G}
 \sum_a  i\bar{\Psi}_a {\cal D}(c_a) \Psi_a ~ , \cr
   {\cal D}(c_a)
    &\!\!\!=\!\!\!&
      \Gamma^A e_A^M
      (\partial_M +{1\over 8}\omega_{MBC}
       [\Gamma^B, \Gamma^C]
       -ig_A A_M
       -ig_B Q_{Xa} B_M)
     -c_a \sigma'(y) ~ ,
     \label{sfermi}
\eea
where the Dirac conjugate is given by 
$\bar{\Psi} =i\Psi^\dag \Gamma^0$
and $\Gamma^\mu$ matrices are given by
\bea
  \Gamma^\mu = \left(\begin{array}{cc}
     & \sigma^\mu \\
     \bar{\sigma}^\mu & \\
     \end{array}\right) ,\quad
 \Gamma^5 =\left(\begin{array}{cc}
    1 & \\
     & -1 \\
     \end{array}\right) ,
   \quad
   \sigma^\mu= (1,\vec{\sigma}) , \quad
   \bar{\sigma}^\mu=(-1,\vec{\sigma}) .
\eea
The non-vanishing spin connection is
$\omega_{\mu m5} =-\sigma' e^{-\sigma}\delta_{\mu m}$, where 
$\delta_{\mu m}$ denotes a vierbein in
the 4D Minkowski spacetime.
The $c_a$ term in Eq.~(\ref{sfermi}) gives
a bulk kink mass, where
$\sigma'(y)=k\epsilon(y)$ is a periodic
step function with a magnitude $k$.
The dimensionless parameter $c_a$ plays an important 
role in the RS warped spacetime.
The orbifold boundary conditions are given by
\bea
  \Psi_a (x,y_j-y) &\!\!\!=\!\!\!& P_j \Gamma^5 \Psi_a (x,y_j+y) .
\eea
With $P_j$ in Eq.~(\ref{pj}),  the first four component of
$\Psi_a$ are even under parity for the 4D left-handed 
$(\Gamma^5=-1)$ components, whereas the fifth component of $\Psi_a$ is even
for the 4D right-handed component.
An $SO(5)$ vector $\Psi$ can be expressed
as the sum of
$({1\over 2},{1\over 2})$ representation and a singlet $(0,0)$ of the subgroup 
$SU(2)_L\times SU(2)_R$.
The $({1\over 2},{1\over 2})$ representation is expressed as
\beqn
\hat{\psi}&\!\!\!=\!\!\!&
\begin{pmatrix}
      \hat{\psi}_{11} & \hat{\psi}_{12} \cr
      \hat{\psi}_{21} & \hat{\psi}_{22} 
\end{pmatrix} 
= \frac{1}{\sqrt{2}}   (\psi_4 +i\vec{\psi}\cdot\vec{\sigma})    i\sigma_2 \cr
\noalign{\kern 10pt}
&\!\!\!=\!\!\!& - \frac{1}{\sqrt{2}} 
\begin{pmatrix}
     \psi_2 +i\psi_1 & -(\psi_4 +i\psi_3) \\
     \psi_4 -i\psi_3 &  \psi_2 -i\psi_1 \\
\end{pmatrix} ~.
\label{fermion1}
\eeqn
$\psi_5$ is a singlet $(0,0)$.
The quarks in the third generation, for instance, 
are composed of bulk Dirac fermions in the  $SO(5)$ vectorial representation
\bea
\Psi_1
 &\!\!\!=\!\!\!&
 \left[Q_1 =\left(\begin{array}{c}
     T \\
     B \\
     \end{array}\right) ,~
    q=\left(\begin{array}{c}
     t \\
     b \\
     \end{array}\right) ,~
   t' \right] ,  \cr
\noalign{\kern 5pt}
 \Psi_2
 &\!\!\!=\!\!\!&
 \left[Q_2 =\left(\begin{array}{c}
     U \\
     D \\
     \end{array}\right) ,~
    Q_3=\left(\begin{array}{c}
     X \\
     Y \\
     \end{array}\right) ,~
   b' \right] ,  
\label{fermion2}
\eea
and  right-handed brane fermions of  the $({1\over 2},0)$
representation in $SU(2)_L \times SU(2)_R$
\bea
   \hat{\chi}_{1R} =
    \left(\begin{array}{c}
       \hat{T}_R \\
       \hat{B}_R \\
       \end{array}\right) ,  \qquad
   \hat{\chi}_{2R} =
    \left(\begin{array}{c}
       \hat{U}_R \\
       \hat{D}_R \\
       \end{array}\right) , \qquad
\hat{\chi}_{3R} =
    \left(\begin{array}{c}
       \hat{X}_R \\
       \hat{Y}_R \\
       \end{array}\right) .    
\eea   
For brane fermions,
the hypercharge $Y/2$ is equal to 
the $U(1)_X$ charge, $Q_X$.
Leptons in the third generation (with massless $\nu_\tau$) are
composed of bulk Dirac fermions in the  $SO(5)$ vectorial
representation
\bea
   \Psi_3 =\left[
   \ell =\left(\begin{array}{c}
      \nu_\tau \\
      \tau \\
      \end{array}\right) ,~
    L_1 =\left(\begin{array}{c}
      L_{1X} \\
      L_{1Y} \\
      \end{array}\right) ,~
      \tau'  \right] ,
\eea
and  right-handed brane fermions of
the $({1\over 2},0)$ representation in $SU(2)_L \times SU(2)_R$
\bea
   \hat{\chi}_{1R}^\ell =\left(\begin{array}{c}
     \hat{L}_{1XR} \\
     \hat{L}_{1YR} \\
     \end{array}\right) .
\eea
%
The assignment of charges for quarks and leptons
is tabulated in Table~\ref{tab:charges},
where the $U(1)_X$ charge, the hypercharge and
the electric charge are denoted as
$Q_X$, $Y/2=T^{3_R} +Q_X$ and 
$Q_E=T^{3_L}+T^{3_R}+Q_X$, respectively. 

\begin{table}[h]
\begin{center}
\caption{Content of bulk and brane fermions in the third generation.  
The brane fermion $\hat\chi_{2R}$ couples to both $q$ and $Q_2$
so that it may be placed in the column of $\Psi_1$ instead of $\Psi_2$.
 \label{tab:charges}}
\begin{tabular}{|ccccc|ccc|}
\noalign{\kern 10pt}
 \hline 
 bulk  & $Q_X$ & & $\onehalf  Y$ & $Q_E$ 
   & on the brane & $Q_X,  \onehalf Y$ & $Q_E$ \\
 \hline
 \noalign{\kern 2pt}
 \hline
 $\Psi_1$ & ${2/ 3}$ &
   $Q_1=\left(\begin{array}{c}
     T \\
     B \\
     \end{array}\right)$
  & ${7/ 6}$ &
 $\begin{array}{c}
   {5/ 3} \\
   {2/ 3} \\
   \end{array}$ 
  &
 $\hat{\chi}_{1R}=\left(\begin{array}{c}
     \hat{T}_R \\
     \hat{B}_R \\
     \end{array}\right)$
  & ${7/ 6}$ &
 $\begin{array}{c}
   {5/ 3} \\
   {2/ 3} \\
   \end{array}$ 
\\   
  &&
 $q=\left(\begin{array}{c}
  t \\
  b \\
  \end{array}\right)$ &
 ${1/ 6}$&
   $\begin{array}{c}
    {2/ 3} \\
    -{1/ 3} \\
  \end{array}$ &&& \\
  &&
  $t'$ & ${2/ 3}$ &
  ${2/ 3}$ &&& \\
 \hline 
 $\Psi_2$ & $-{1/ 3}$ &
   $Q_2=\left(\begin{array}{c}
     U \\
     D \\
     \end{array}\right)$
  & ${1/ 6}$ &
 $\begin{array}{c}
   {2/ 3} \\
   -{1/ 3} \\
   \end{array}$ 
 &
 $\hat{\chi}_{2R}=\left(\begin{array}{c}
  \hat{U}_R \\
  \hat{D}_R \\
  \end{array}\right)$ &
 ${1/ 6}$&
   $\begin{array}{c}
    {2/ 3} \\
    -{1/ 3} \\
  \end{array}$ 
\\
  &&
 $Q_3=\left(\begin{array}{c}
  X \\
  Y \\
  \end{array}\right)$ &
 $-{5/ 6}$&
   $\begin{array}{c}
    -{1/ 3} \\
    -{4/ 3} \\
  \end{array}$ 
 &  
$\hat{\chi}_{3R}=\left(\begin{array}{c}
     \hat{X}_R \\
     \hat{Y}_R \\
     \end{array}\right)$
  & $-{5/ 6}$ &
 $\begin{array}{c}
   -{1/ 3} \\
   -{4/ 3} \\
   \end{array}$ 
\\
 && $b'$ & $-{1/ 3}$ &
  $-{1/ 3}$ &&& \\
  \hline
 $\Psi_3$ & $-1$ &
   $\ell=\left(\begin{array}{c}
     \nu_\tau \\
     \tau \\
     \end{array}\right)$
  & $-{1/ 2}$ &
 $\begin{array}{c}
   0 \\
   -1 \\
   \end{array}$ &&& \\
  &&
 $L_1=\left(\begin{array}{c}
  L_{1X} \\
  L_{1Y } \\
  \end{array}\right)$ &
 $-{3/ 2}$&
   $\begin{array}{c}
    -1 \\
    -2 \\
  \end{array}$ 
 &
 $\hat{\chi}_{1R}^\ell=\left(\begin{array}{c}
  \hat{L}_{1XR} \\
  \hat{L}_{1YR} \\
  \end{array}\right)$ &
 $-{3/ 2}$&
   $\begin{array}{c}
    -1 \\
    -2 \\
  \end{array}$
\\
 && $\tau'$ & $-1$ &
  $-1$ &&& \\
\hline

\noalign{\kern 2pt}
 \hline
 $\Psi_4$ & $0$ &
   $L_2=\left(\begin{array}{c}
     L_{2X} \\
     L_{2Y} \\
     \end{array}\right)$
  & ${1/2}$ &
 $\begin{array}{c}
   1\\
   0 \\
   \end{array}$ 
  &
 $\hat{\chi}_{2R}^\ell =\left(\begin{array}{c}
     \hat{L}_{2XR} \\
     \hat{L}_{2YR}  \\
     \end{array}\right)$
  & $1/2$ &
 $\begin{array}{c}
  1 \\
0 \\
   \end{array}$ 
\\   
  &&
 $L_3=\left(\begin{array}{c}
     L_{3X} \\
     L_{3Y} \\
  \end{array}\right)$ &
 ${-1/2}$&
   $\begin{array}{c}
   0 \\
    -1 \\
  \end{array}$ 
  &
  $\hat{\chi}_{3R}^\ell =\left(\begin{array}{c}
     \hat{L}_{3XR} \\
     \hat{L}_{3YR}  \\
     \end{array}\right)$
  & $-1/2$ &
 $\begin{array}{c}
0 \\
-1 \\
   \end{array}$ 
\\
  &&
  $\nu_\tau'$ & $0$ &
  $0$ &&& \\
 \hline 
\end{tabular}
\end{center}
\end{table}  

The component $(t-B)/\sqrt{2}$ mixes with $t'$ through
the nonvanishing Wilson line phase $\theta_H$.  
Similarly, $(X-D)/\sqrt{2}$ and $(L_{1X}-\tau)/\sqrt{2}$ mix with
$b'$ and $\tau'$, respectively. With $\theta_H$ alone, there remain 
many unwanted massless modes of fermions. To make them heavy 
we introduce  right-handed fermions $\hat \chi_{\alpha R}$
and $\hat{\chi}_{\alpha R}^\ell$  
in the $(\onehalf, 0)$ 
representation of $SU(2)_L \times SU(2)_R$ localized on the Planck
brane at $y=0$ as tabulated in Table \ref{tab:charges}.

The brane fermions $\hat \chi_{\alpha R}$ and $\hat \chi_{\alpha R}^\ell$
couple to the corresponding bulk fermions
and the brane scalar 
$\Phi$ in (\ref{cov})
without spoiling the $SO(4)$ symmetry on the Planck brane
through Yukawa couplings 
such as $\tilde{y}\hat{\chi}_{2R}^\dag(Q_1 q) \Phi$ and
$y_1 \hat{\chi}_{1R}^\dag (Q_1 q) \tilde{\Phi}$, etc.,
where $(Q_1 q)$ is cast in a 2-by-2 matrix in the $({1\over 2},{1\over 2})$
representation of $SU(2)_L \times SU(2)_R$.
The conjugate $SU(2)_R$-doublet 
and the Yukawa coupling constants are denoted as
$\tilde{\Phi}=i\sigma^2 \Phi^*$ and
$\tilde{y}$, $y_1$, respectively.

After the scalar field $\Phi$ develops a vacuum expectation value,  
the general brane action for $\hat \chi_{\alpha R}$ and $\hat \chi_{\alpha R}^\ell$
becomes
\beqn
&&\hskip -1cm
S_{\rm Pl. brane}^{\rm fermion} =\int d^5x \sqrt{-G} ~  i \delta(y) \cr
\noalign{\kern 5pt}
&&\hskip -.7cm
\times \bigg\{ 
   \sum_{\alpha=1}^3  \Big[ \hat{\chi}_{\alpha R}^\dag
   \bar{\sigma}^\mu   D_\mu \hat{\chi}_{\alpha R}
    - \mu_\alpha (\hat{\chi}_{\alpha R}^\dag Q_{\alpha L}
    -Q_{\alpha L}^\dag \hat{\chi}_{\alpha R}) \Big] 
    -\tilde{\mu} (\hat{\chi}_{2R}^\dag q_L   -q_L^\dag \hat{\chi}_{2R})  \cr
\noalign{\kern 5pt}
&&\hskip -.4cm
+  \sum_{\alpha=1}^3  \Big[ \hat{\chi}_{\alpha R}^{\ell \dag}
   \bar{\sigma}^\mu   D_\mu \hat{\chi}_{\alpha R}^\ell 
    - \mu_\alpha^\ell  (\hat{\chi}_{\alpha R}^{\ell \dag} L_{\alpha L}
    -L_{\alpha L}^\dag \hat{\chi}_{\alpha R}^\ell ) \Big] 
    -\tilde{\mu}^\ell  (\hat{\chi}_{3R}^{\ell \dag} \ell_L   -\ell_L^\dag \hat{\chi}_{3R}^\ell)  
\bigg\} ~,
\label{brane1}
\eeqn
where $D_\mu$ in the kinetic term has the same form as in Eq.~(\ref{cov}) 
with $A_\mu^{a_R}T^{a_R}$ replaced by $A_\mu^{a_L}T^{a_L}$.
The $\mu$ terms mix bulk left-handed fermions and brane right-handed fermions.
The couplings $\mu_\alpha, \tilde \mu, \mu_\alpha^\ell$ and $\tilde \mu^\ell$ have
the dimension $M^{1/2}$. 
As we shall see below, only modest conditions are necessary to be satisfied 
for the values of  $\mu$'s to get the desired low energy mass spectrum.  
In the case of massless neutrinos,  $\Psi_4$, $\hat{\chi}_{2R}^\ell$
and $\hat{\chi}_{3R}^\ell$ are unnecessary.

\section{Anomaly cancellation} 

In the previous section  the brane fermions 
$\hat \chi_{\alpha R}$ and $\hat \chi_{\alpha R}^\ell$ are introduced
to get the low energy spectrum of quarks and leptons.
We would like to show that they are necessary \myred{for the 
cancellation of 4D anomalies} with respect to the 
$SU(2)_L\times SU(2)_R \times U(1)_X$ gauge symmetry.  
From this viewpoint the presence of the brane 
fermions is expected to be a physical consequence resulting from a more
fundamental theory.

Let us first consider the case of massless neutrinos
in which there are the multiplets $\Psi_1$, $\Psi_2$, $\hat\chi_{1R}$, 
$\hat\chi_{2R}$, and $\hat\chi_{3R}$ in the quark sector, and
 $\Psi_3$ and  $\hat\chi_{1R}^\ell$ in the lepton sector.
We need to check the cancellation of \myred{4D anomalies} for triangle diagrams
in the gauge group $SU(2)_L \times SU(2)_R \times U(1)_X (\times SU(3)_c) $.
We recall that the first four components of  left-handed $\Psi_a$ and
the fifth component of  right-handed $\Psi_a$ have chiral zero modes.
Brane fermions are all right-handed.  
\myred{The cancellation is achieved irrespective of the values of
the brane masses $\mu_\alpha$, $\tilde \mu$, $\mu_\alpha^\ell$, 
and $\tilde \mu^\ell$.}

From the assignment of charges in  Table~\ref{tab:charges},  
the anomaly cancellation for $SU(2)_L\times U(1)_Y$ is manifest.
For three $U(1)_X$ bosons,  anomalous terms for quarks for each color
are proportional to
\bea
 &&  \left({2\over 3}\right)^3 \times (-4 +1)
   +\left(-{1\over 3}\right)^3
    \times (-4 +1)
\nonumber
\\
  &&
  +\left({7\over 6}\right)^3\times 
     2
   +\left({1\over 6}\right)^3 \times 2
   +\left(-{5\over 6}\right)^3 \times 2
  =  {5\over 4} ,
\eea    
and the coefficient for leptons is
\bea
   (-1)^3 \times (-4+1) +\left(
  -{3\over 2}\right)^3 \times 2
   =-{15\over 4}.
\eea
For three $U(1)_X$ bosons, the sum of contributions vanishes,
\bea
 3\times {5\over 4} -{15\over 4} =0 .
\eea
For one $U(1)_X$ boson, the coefficient for quarks 
for each color is
\bea
   {2\over 3} \times (-4 +1)
  +\left(-{1\over 3}\right) \times (-4+1)
  +{7\over 6}\times 2 
   +{1\over 6}\times 2 
   +\left(-{5\over 6}\right)\times 2
  =0 ,
   \label{u1x}
\eea 
and the coefficient for leptons is
\bea
   -1 \times (-4+1) +\left(-{3\over 2}\right) \times 2
  =0 .
\eea
From Eq.~(\ref{u1x}), the coefficient for one $U(1)_X$ boson with
two $SU(3)_c$ bosons is also vanishing.  
For one $U(1)_X$ boson with two  $SU(2)_L$ bosons, the coefficient is
\bea
  {2\over 3}\times (-2) + \left(-{1\over 3}\right)
  \times (-2) 
   +{7\over 6}\times 1
  +{1\over 6}\times 1  
  +\left(-{5\over 6}\right) \times 1 
  =-{1\over 6} ,
\eea
for quarks for each color   and 
\bea
   -1 \times (-2) + \left(-{3\over 2}\right) \times 1
   ={1\over 2} ,
\eea
for leptons.  The sum of the contributions  vanishes.  
Finally for one $U(1)_X$ boson with two $SU(2)_R$ bosons, the coefficient is
\bea
   {2\over 3} \times (-2) 
 -{1\over 3} \times (-2)
  =-{2\over 3} ,
\eea
for quarks for each color and 
\bea
  -1 \times (-2) =2,
\eea
for leptons.  The sum vanishes.
We have confirmed the anomaly cancellation for $\Psi_i$,
$\hat{\chi}_{iR}$, $(i=1,2,3)$ and $\hat{\chi}_{1R}^\ell$. 

In the case of massive neutrinos the set of  additional multiplets,
 $(\Psi_4,  \hat{\chi}_{2R}^\ell,  \hat{\chi}_{3R}^\ell)$ must be included.
 Amazingly this set of the multiplets does not generate any anomaly.
 Indeed, for three $U(1)_X$, anomalous terms for 
$(\Psi_4, \hat{\chi}_{2R}^\ell, \hat{\chi}_{3R}^\ell)$
are proportional to 
\bea
 0\times (-4+1) + \left({1\over 2}\right)^3\times 2
+\left(-{1\over 2}\right)^3 \times 2 =0 .
\eea
For one $U(1)_X$, the coefficient is
\bea
  0\times (-4+1) +\left({1\over 2}\right)\times 2
   +\left(-{1\over 2}\right)\times 2= 0 .
\eea
From this equation, the coefficient for one $U(1)_X$ boson with two
$SU(3)_c$ bosons is also vanishing.
For one $U(1)_X$ boson with two $SU(2)_L$ bosons, the
coefficient is 
\bea
  0\times (-2) +\left({1\over 2}\right) \times 1
   +\left(-{1\over 2}\right) \times 1 =0 .
\eea
Finally for one $U(1)_X$ boson with two 
$SU(2)_R$ bosons, the coefficient is
\bea
  0 \times (-2) = 0.
\eea
Thus \myred{the cancellation of the 4D anomalies} has been confirmed  even when  
$(\Psi_4,  \hat{\chi}_{2R}^\ell,  \hat{\chi}_{3R}^\ell)$
is also included.

\myred{We remark that there remain 5D anomalies.  As shown in 
refs.\ \cite{Arkani-Hamed1} - \cite{Hirayama}, the bulk fermions give rise
to an anomaly to a 5D current $J^M_{\rm 5D}(x,y)$.  
The anomaly associated with bulk fermions takes the form 
\myblue{${J^M_{\rm 5D~bulk}}_{;M} = a \{ \delta(y) + \delta(y-L) \} 
F_{\mu\nu} \tilde F^{\mu\nu}/ \sqrt{-G}$}.  
With the contributions from brane
fermions incorporated, it becomes 
\myblue{${J^M_{\rm 5D}}_{;M} = a \{ - \delta(y) + \delta(y-L) \} 
F_{\mu\nu} \tilde F^{\mu\nu}  / \sqrt{-G}$} and the cancellation 
of the 4D anomaly is ensured; $\dd_\mu J^\mu_{\rm 4D}(x) =0$ 
\myblue{with $J^\mu_{\rm 4D}(x) = \int dy \sqrt{ -G} \, J^\mu_{\rm 5D}(x,y)$}.  
The remaining 5D anomalies need be cancelled, which may be achieved, 
for instance, by  introducing Chern-Simons terms.  We leave more detailed 
analysis  for future investigation.}

\section{Spectrum and mode function profiles}

In identifying the spectrum of particles and their wave functions,
the conformal coordinate $z=e^{\sigma(y)}$ for the fifth dimension is useful, 
with which the metric becomes
\bea
   ds^2 &\!\!\!=\!\!\!& {1\over z^2} \left\{ \eta_{\mu\nu} dx^\mu dx^\nu + {dz^2\over k^2}\right\} .
\eea
The fundamental region $0\leq y\leq L$ is  mapped to $1\leq z \leq z_L=e^{kL}$.
In the bulk region $0<y<L$, one has  $\partial_y =kz \partial_z$,
$A_y =kz A_z$, $B_y = kz B_z$.
To find the gauge couplings of quarks and leptons in four dimensions one needs
their wave functions in the fifth dimension.

\subsection{Gauge bosons}

The $SO(5)$ gauge fields are split into classical and quantum parts
$A_M =A_M^c + A_M^q$, where
$A_\mu^c =0$ and $A_y^c = (dz/dy) A_z^c  =kz A_z^c$. 
With the gauge-fixing functional 
\bea
    f_{\textrm{\scriptsize gf}}^{(A)} = 
    z^2 \left\{ \eta^{\mu\nu} {\cal D}_\mu^c A_\nu^q
    +\xi k^2 z {\cal D}_z^c \left({1\over z} A_z^q\right) \right\} ,
\eea
the quadratic part of the action for the $SO(5)$ gauge fields is given by 
\bea
   S_{\textrm{\scriptsize bulk 2}}^{%
  \textrm{\scriptsize gauge}} =
   \int d^4 x {dz\over kz}
  \left[ \textrm{tr}\left\{
   \eta^{\mu\nu} A_\mu^q (\Box + k^2 {\cal P}_4) A_\nu^q
   +k^2 A_z^q (\Box +k^2 {\cal P}_z)A_z^q 
   \right\} \right] ,
\eea
for $\xi=1$. Here $A_\mu^c=0$ have been taken.
The differential operators are defined by 
$\Box = \eta^{\mu\nu}\partial_\mu \partial_\nu$,
${\cal P}_4= z{\cal D}_z^c ({1/ z}) {\cal D}_z^c$,
${\cal P}_z = {\cal D}_z^c z {\cal D}_z^c ({1/ z})$ 
where ${\cal D}_M^c A_N^q = \partial_M A_N^q - ig_A [ A_M^c, A_N^q]$.   
The linearized equations of motion are
\bea
   \Box A_\mu^q + k^2 z {\cal D}_z^c {1\over z} 
  {\cal D}_z^c A_\mu^q 
   =0 ,
\qquad
   \Box A_z^q + k^2 {\cal D}_z^c z 
  {\cal D}_z^c {1\over z} A_z^q 
   =0 .
 \label{eom1}
\eea

The $SO(4)$ vector $A_y^{\hat{a}}$,  which forms
an $SU(2)_L$-doublet
$\Phi_H^t =(
A_y^{\hat{2}} +i A_y^{\hat{1}},
A_y^{\hat{4}} -i A_y^{\hat{3}})$, 
has zero modes.
One can utilize the residual symmetry such that  
the zero mode of $A_y^{\hat{4}}$ yield a nonzero vacuum expectation value 
$\langle A_y^{\hat{a}}\rangle =v \, \delta^{a4}$.
The Wilson line phase $\theta_H$ is given by
$\exp \{{i\over 2}\theta_H (2\sqrt{2}T^{\hat{4}})\}
=\exp\{ig_A\int_1^{z_L} dz \langle A_z\rangle\}$
so that 
$\theta_H ={1\over 2}
g_A v \sqrt{(z_L^2 -1)/k}$~\cite{HM}.
By a large gauge transformation which maintains the orbifold boundary conditions
$\theta_H$ is shifted to $\theta_H + 2\pi$.
The gauge invariance of the theory implies that physics is periodic in $\theta_H$
with a period $2\pi$.
By a gauge transformation 
a new basis can be taken in which the background 
field vanishes, $\tilde{A}_z^c=0$.~\cite{HS2, Falkowski1}
The new gauge is called the twisted gauge as the boundary conditions are
twisted.  The new gauge potentials  are related to the original ones by
$\tilde{A}_M = \Omega A_M^q \Omega^{-1}$, $\tilde{B}_M =B_M^q$ where
$\Omega(z) = \exp \{ i\theta(z) \sqrt{2}T^{\hat{4}}\}$ 
and $\theta(z) =\theta_H(z_L^2 -z^2)/(z_L^2 -1)$. 
The equations of motion (\ref{eom1}) become
 \bea
     \Box \tilde{A}_\mu + k^2 \left(\partial_z^2 -{1\over z} \partial_z\right) 
   \tilde{A}_\mu =0 ,
\quad
    \Box \tilde{A}_z + k^2 \left(\partial_z^2 -{1\over z}\partial_z +{1\over z^2}\right)
    \tilde{A}_z =0 .
\eea
$\tilde{B}_M$ satisfies the same equations as $\tilde{A}_M$.

The four-dimensional components of the $SO(5)$ and $U(1)_X$ gauge bosons 
contain $W$ and $Z$ bosons and photon as
\bea
   \tilde{A}_\mu(x,z)
     &\!\!\!=\!\!\!&
   W_\mu \left\{
      h_W^L T^{-_L} + h_W^R T^{-_R} + h_W^\wedge T^{\hat{-}} \right\}
   +   
    W_\mu^\dag \left\{
      h_W^L T^{+_L} + h_W^R T^{+_R} + h_W^\wedge T^{\hat{+}}
        \right\}
\nonumber
\\
  &&\qquad
   +Z_\mu \left\{
     h_Z^L T^{3_L} +h_Z^R T^{3_R} + h_Z^\wedge T^{\hat{3}} 
  \right\}
  +A^\gamma_\mu
      h_\gamma\left\{ T^{3_L} +T^{3_R}\right\}  + \cdots , 
\nonumber
\\
\noalign{\kern 10pt}
  \tilde{B}_\mu(x,z)&\!\!\!=\!\!\!&
     Z_\mu h_Z^B +A_\mu^\gamma h_\gamma^B  + \cdots .
\eea
The wave functions $h_W^L(z)$, $h_W^R(z)$
and $h_W^\wedge(z)$ for $W_\mu(x)$,
$h_Z^L(z)$, $h_Z^R(z)$, $h_Z^\wedge(z)$ and
$h_Z^B(z)$ for $Z(x)$ and
$h_\gamma(z)$ and $h_\gamma^B(z)$ for $A_\gamma(x)$ 
satisfy their own equations of motion and boundary conditions. 
The boundary conditions at $z=1$ is summarized in Appendix~\ref{ap:bcg}.
\ignore{The lowest mass modes for $W_\mu(x)$, $Z_\mu(x)$ and $A_\gamma(x)$
are $W$ and $Z$ bosons and photon, respectively.}

\subsubsection*{(i) $W$ boson tower}

The wave functions of the KK tower of $W_\mu(x)$ are given by
\beqn
\begin{pmatrix} h_W^L\cr h_W^R \end{pmatrix} &\!\!\!=\!\!\!&
\frac{1 \pm \cos\theta_H}{2} \,  N_W(z;\lambda) ~, \cr
\noalign{\kern 5pt}
h_W^\wedge ~ &\!\!\!=\!\!\!&
        -\frac{\sin\theta_H}{\sqrt{2}} \, D_W(z;\lambda) ~.
\label{hw3}
\eeqn
Here  $N_W(z;\lambda) =2aC(z;\lambda)$
and $D_W(z;\lambda)=2a(C(1;\lambda)/S(1;\lambda)) S(z;\lambda)$
where the $C$ and $S$ functions are defined as~\cite{HOOS, Falkowski1}
\bea
&&\hskip -1cm 
C(z;\lambda) =
      {\pi\over 2}\lambda z z_L F_{1,0}(\lambda z,
        \lambda z_L) ~,
\quad
   C'(z;\lambda) =
      {\pi\over 2}\lambda^2 z z_L F_{0,0}(\lambda z,
        \lambda z_L) ~ ,
\cr
\noalign{\kern 5pt}
&&\hskip -1cm 
S(z;\lambda) =
      -{\pi\over 2}\lambda z  F_{1,1}(\lambda z,
        \lambda z_L) ~ ,
\quad
   S'(z;\lambda) =
      -{\pi\over 2}\lambda^2 z F_{0,1}(\lambda z,
        \lambda z_L) ~, 
        \cr
\noalign{\kern 5pt}
&&\hskip -1cm 
F_{\alpha,\beta}(u,v) =
    J_\alpha (u) Y_\beta(v)   -Y_\alpha(u) J_\beta(v) ~.
\eea
Here the prime denotes a derivative;  $C'=dC/dz$ etc..
A relation  $CS' -SC' =\lambda z$ holds. 
From the normalization condition 
$\int_1^{z_L} dz (kz)^{-1} \left\{ 
      (h_W^L)^2+(h_W^R)^2+(h_W^\wedge)^2\right\}  =1$
the coefficient $a$ is determined to be
\beeq
a^{-2} =  \int_1^{z_L}  {dz\over kz}  2 \bigg\{ 2(C(z;\lambda))^2
    -\sin^2\theta_H
  \bigg[  (C(z;\lambda))^2
    -\bigg({C(1;\lambda)\over S(1;\lambda)}\bigg)^2
    (S(z;\lambda))^2 \bigg]  \bigg\}  ~.
\eneq

\begin{figure}[htb]
\begin{center}
\includegraphics[height=5.cm]{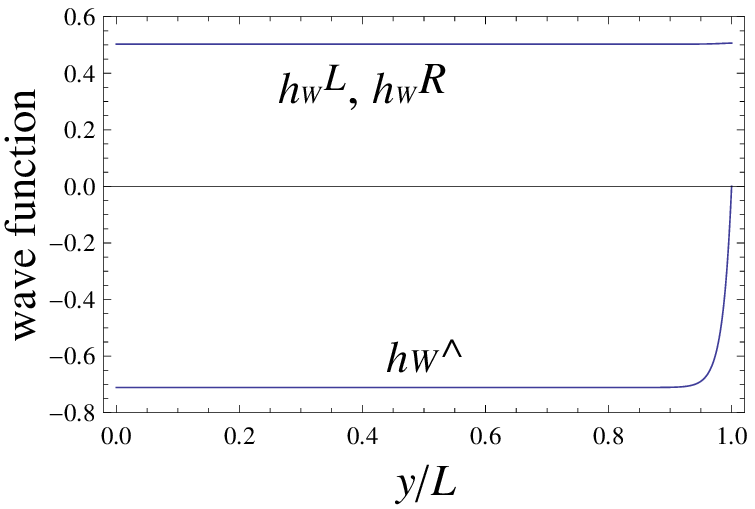}
~~
\includegraphics[height=5.cm]{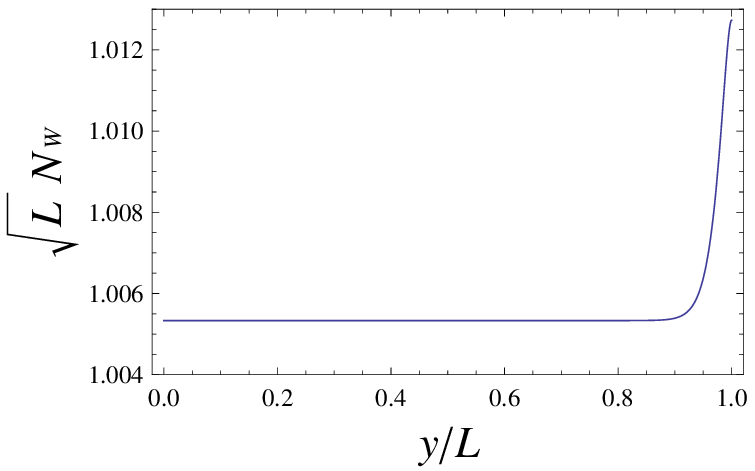} 
\caption{The $y$-dependence of the wave functions $h_W^L$, $h_W^R$ and 
 $h_W^\wedge$ of the $W$ boson.  
Here $z_L =10^{15}$, $k=4.7\times 10^{17}$~GeV, $m_W=80.398$~GeV,
$\theta_H=\pi/2$.  
\label{fig:hw}}
\end{center}
\end{figure}

The mass spectrum $k \lambda$ for $W_\mu(x)$ and its KK tower is determined by
\bea
   2S(1;\lambda)C'(1;\lambda) 
  + \lambda \sin^2\theta_H  =0 ~ .
 \label{bcw}
\eea
The mass $m_W$ of the $W$ boson, the lightest mode, is given by
\bea
  m_W = k\lambda_W
  \approx \sqrt{k\over L} e^{-kL} |\sin\theta_H|
  \approx {m_\textrm{\scriptsize KK}\over \pi \sqrt{kL}}
  |\sin\theta_H|  ~ .
\eea
Behavior of the wave functions with respect to
$y/L$ is shown in Fig.~\ref{fig:hw}.
The wave function of the $W$ boson is flat in most of the region in the bulk in the 
RS space, which should be contrasted to the case of models in a flat spacetime.
In the right figure,  the detailed $y$-dependence of $\sqrt{L}N_W$ 
\myred{is}
depicted.  
In the  region where the wave function is flat, 
$\sqrt{L}N_W \simeq1.00533$.
$\sqrt{L} D_W\simeq 1-z^2/z_L^2$ for $z/z_L \ll 1$.

\subsubsection*{(ii) Photon tower}

The wave functions of the $A_\mu^\gamma$ bosons in the photon tower
are given by
\bea
   h_\gamma = c s_\phi C(z;\lambda) ,
   \qquad
   h_\gamma^B = c c_\phi C(z;\lambda) 
   \label{hg}
\eea
where the coefficient $c$ is determined by the normalization 
$\int_1^{z_L} dz \, (kz)^{-1}  (2(h_\gamma)^2+ (h_\gamma^B)^2)=1$.
The mass spectrum is determined by 
\bea
   C'(1;\lambda) =0 .
   \label{agm}
\eea
The lowest mass mode is the massless  photon.
Its wave functions  are constant; 
\bea
   h_\gamma =
    {s_\phi\over\sqrt{(1+s_\phi^2)L}} ,
\qquad 
  h_\gamma^B
    ={c_\phi\over \sqrt{(1+s_\phi^2)L}} .
\eea     
It follows from
$g_A h_\gamma(T^{3_L} +T^{3_R}) +g_B Q_X h_\gamma^B = 
Q_E g_A g_B/\sqrt{(g_A^2 +2g_B^2)L}$ that the 4D gauge couplings for weak and electromagnetic interactions are
\bea
   \myred{\bar{g}_A} ={g_A\over \sqrt{L}} ~, \qquad
   e = {g_A g_B\over \sqrt{(g_A^2 +2g_B^2)L}} ~,
 \label{4Dcoupling1}
\eea 
which lead to $e=\myred{\bar{g}_A}
s_\phi /\sqrt{1+s_\phi^2}$
and $s_\phi^2 =\tan^2\theta_W$ where $\theta_W$ is the weak mixing angle.

\subsubsection*{(iii) $Z$ boson tower}

The wave functions of the bosons in the $Z$ boson tower are
\beqn
\begin{pmatrix} h_Z^L \cr h_Z^R \end{pmatrix} &\!\!\!=\!\!\!&
\frac{ c_\phi^2 \pm \cos \theta_H (1+s_\phi^2)}{2\sqrt{1+s_\phi^2}}
   ~ N_Z(z;\lambda) ~ , \cr
\noalign{\kern 5pt}
h_Z^\wedge ~~ &\!\!\!=\!\!\!&
\myred{
     - \frac{1}{ \sqrt{2}} 
     \sin\theta_H  \sqrt{1+s_\phi^2}
     ~ D_Z(z;\lambda) } ~, \cr
\noalign{\kern 8pt}
h_Z^B ~~ &\!\!\!=\!\!\!&  
\myred{
- \frac{s_\phi c_\phi}{\sqrt{1+s_\phi^2}} ~ N_Z(z;\lambda) } ~.
\label{hz}
\eeqn
Here $N_Z =2b\sqrt{1+s_\phi^2} C(z;\lambda)$
and $D_Z=2b\sqrt{1+s_\phi^2} (C(1;\lambda)/S(1;\lambda)) S(z;\lambda)$.
From the normalization condition 
$\int_1^{z_L}dz \,  (kz)^{-1} \{(h_Z^L)^2 + (h_Z^R)^2 + (h_Z^\wedge)^2
 +(h_Z^B)^2 \} =1$ the coefficient $b$ is determined to be
\beqn
&&\hskip -1cm 
b^{-2} =  \int_1^{z_L} \frac{dz}{kz}   \,  2(1+s_\phi^2)^2 \cr
\noalign{\kern 10pt}
&&\hskip -.5cm
\times  \Bigg\{  \frac{2}{1+s_\phi^2} \,  (C(z;\lambda))^2
-\sin^2 \theta_H \bigg[ C(z;\lambda) ^2  
-\bigg( \frac{C(1;\lambda)}{S(1;\lambda)} \bigg)^2
  S(z;\lambda)^2 \bigg] \Bigg\} ~.
\eeqn

\begin{figure}[htb]
\begin{center}
\includegraphics[height=5cm]{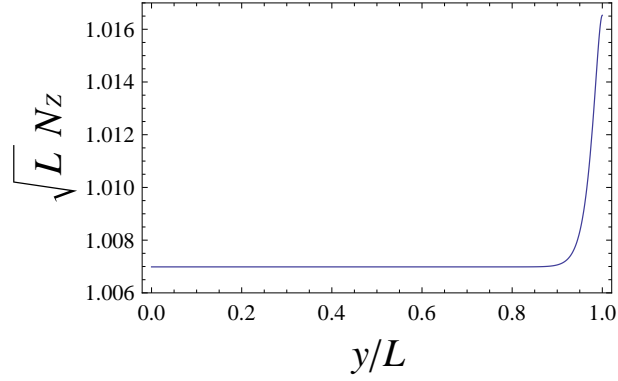}
\caption{The $y$-dependence of $\sqrt{L} N_Z$ for the $Z$ boson. 
Here $z_L =10^{15}$, $k=4.7\times 10^{17}$~GeV,
$m_Z=91.1876$~GeV an d $\theta_H = \pi/2$. 
\label{fig:hz}}
\end{center}
\end{figure}

The mass spectrum of the  $Z$ tower is determined by
\bea
   2S(1;\lambda) C'(1;\lambda)
     +\lambda(1+s_\phi^2) \sin^2\theta_H =0  ~.
     \label{zm}
\eea
The mass of  the lightest mode, the $Z$ boson, is given by
\bea
    m_Z \approx {m_W\over \cos \theta_W} ~ .
\eea
The profile of the wave functions of the  $Z$ boson
is similar to that of the $W$ boson up to overall factors. 
The dominant contribution to the normalization comes
from $N_Z = 2b\sqrt{1+s_\phi^2}C(z;\lambda)$. 
The figure~\ref{fig:hz} depicts the $y$-dependence of
$\sqrt{L} N_Z$ of the $Z$ boson. 
$\sqrt{L} N_Z \simeq 1.00699$ in the region where the wave function is flat.
$\sqrt{L} D_Z \simeq 1-z^2/z_L^2$ for $z/z_L\ll 1$.

\subsection{Fermions}

In terms of the rescaled fields $\tilde\Psi_a = z^{-2} \Omega(z)  \Psi_a$ 
in the twisted gauge where
\beeq
\Omega(z) = 
\begin{pmatrix} {\bf 1}_3 && \cr
         & \cos \theta(z) & \sin \theta(z) \cr
         & -\sin\theta(z) & \cos\theta(z)      \end{pmatrix}
 ~~,~~ 
 \theta(z) = \frac{z_L^2 - z^2}{z_L^2 -1} \, \theta_H ~~, 
\eneq
the  action for the  fermions in the bulk region becomes
\beqn
&&\hskip -1cm
 S_{\rm bulk}^{\rm fermion} = \sum_a \int d^4x \frac{dz}{k} ~
i {\overline{\tilde\Psi}}_a   \Big\{  \Gamma^\mu (\partial_\mu
         -ig_A \tilde{A}_\mu -ig_B Q_{Xa}  \tilde{B}_\mu)  \cr
\noalign{\kern 10pt}
&&\hskip .8cm
+ \Gamma^5 \sigma'  (\dd_z -ig_A \tilde{A}_z -ig_B Q_{Xa} \tilde{B}_z)
- \frac{c_a}{z} \, \sigma'  \Big\} ~ \tilde{\Psi}_a ~~.
\label{fermiact}
\eeqn
If there were no brane interactions, $\tilde{\Psi}_a$  would obey
\bea
  &&\left\{
     \left( \begin{array}{cc}
        & \sigma \cdot \partial \\
      \bar{\sigma} \cdot \partial & \\
      \end{array}\right)
   -k\left(\begin{array}{cc}
     D_-(c_a) & \\
     & D_+(c_a) \\
   \end{array}\right)\right\}
 \left(\begin{array}{c}
   \tilde{\Psi}_{aR} \\
   \tilde{\Psi}_{aL} \\
   \end{array}\right) =0 ,
\eea
where
$D_{\pm}(c) = \pm (d/dz) + (c/z)$.
Neumann conditions for $\tilde{\Psi}_R$
and $\tilde{\Psi}_L$ are given by
$D_-(c)\tilde{\Psi}_R=0$ and 
$D_+(c)\tilde{\Psi}_L=0$, respectively.
The resulting second order differential equations are
\bea
    \left\{
        \partial^2 - k^2 D_-(c_a) D_+(c_a)
        \right\} \tilde{\Psi}_{aL} =0  ~,
\quad
   \left\{
        \partial^2 - k^2 D_+(c_a) D_-(c_a)
        \right\} \tilde{\Psi}_{aR} =0 ~.
\eea
The basis functions are given by 
\beqn
&&\hskip -1cm
\begin{pmatrix} C_L \cr S_L \end{pmatrix}  (z;\lambda, c)
= \pm \frac{\pi}{2} \lambda\sqrt{zz_L}
   F_{c+{1\over 2},c\mp{1\over 2}}  (\lambda z, \lambda z_L) ~, \cr
\noalign{\kern 10pt}
&&\hskip -1cm  
\begin{pmatrix} C_R \cr S_R \end{pmatrix}  (z;\lambda, c)
= \mp \frac{\pi}{2} \lambda\sqrt{zz_L}
F_{c-{1\over 2},c\pm {1\over 2}} 
 (\lambda z, \lambda z_L)~.
\eeqn
They satisfy
\beqn
&&\hskip -1cm
\big\{ D_+(c) D_-(c) -\lambda^2 \big\} 
\begin{pmatrix} C_R (z) \cr S_R (z) \end{pmatrix}  =0 ~, \cr
\noalign{\kern 5pt}
&&\hskip -1cm
\big\{ D_-(c) D_+(c) -\lambda^2 \big\} 
\begin{pmatrix} C_L (z) \cr S_L (z) \end{pmatrix}  =0 ~, \cr
\noalign{\kern 5pt}
&&\hskip -1cm
S_L(z;\lambda ,-c) =-S_R(z;\lambda,c) ~~,~~ 
C_LC_R -S_L S_R=1 ~.
\eeqn
They obey the boundary conditions that
$C_R=C_L=1$,
$D_- C_R=D_+ C_L =0$,
$S_R=S_L=0$ and
$D_- S_R =D_+ S_L =\lambda$ at $z=z_L$.
Further the operators $D_\pm$ link $L$ and $R$ functions  by
$D_+(C_L,S_L) =\lambda(S_R, C_R)$
and $D_-(C_R,S_R) =\lambda(S_L, C_L)$.

\subsubsection*{(i) Top quark}

The wave functions of fermions have been determined in ref.\ \cite{HK}.
In the quark sector we choose $c_1=c_2=c$.
The top quark component $t(x)$ in four dimensions is contained in the form
\bea
  \left(\begin{array}{c}
  \tilde{U}_L \\
  (\tilde{B}_L \pm \tilde{t}_L)/\sqrt{2} \\
   \tilde{t}_L' \\
  \end{array}\right) (x,z)
   &\!\!\!=\!\!\!&
  \sqrt{k} \left(\begin{array}{c}
  a_U C_L(z;\lambda, c) \\
   a_{B\pm t} C_L(z; \lambda, c) \\
   a_{t'} S_L(z;\lambda, c) \\
  \end{array}\right) t_L(x) ,
\cr
   \left(\begin{array}{c}
  \tilde{U}_R \\
  (\tilde{B}_R \pm \tilde{t}_R)/\sqrt{2} \\
   \tilde{t}_R' \\
  \end{array}\right) (x,z)
   &\!\!\!=\!\!\!&
  \sqrt{k} \left(\begin{array}{c}
  a_U S_R(z;\lambda, c) \\
   a_{B\pm t} S_R(z; \lambda, c) \\
   a_{t'} C_R(z;\lambda, c) \\
  \end{array}\right) t_R(x) .
\eea
For the right-handed components there are tiny mixture of the brane fermions
$\hat \chi_R$.  Their contributions have been shown to be negligibly small.\cite{HK}
The $u$ and $c$ quarks are described in a similar way.
We suppose that the scale of brane masses is much larger than the KK
mass scale; $\mu_1^2, \mu_2^2, \mu_3^2, \tilde\mu^2 \gg m_\KK$.
Then the ratios of the coefficients $a$'s are given by
\bea
   [ a_U , a_{B-t} , a_{t'} ]
   \simeq 
   \left[ -{\sqrt{2}\tilde{\mu}
    \over \mu_2} ,
    -c_H, 
    - \frac{s_H C_L|_{z=1}}{S_L|_{z=1} }
  \right]
    a_{B+t} ~~.
\eea    
Here $c_H =\cos\theta_H$ and $s_H =\sin\theta_H$. 
The coefficient $a_{B+t}$ is determined
by
\bea
  a_{B+t}^{-2}
    =\int_1^{z_L}
     dz
    \left\{
    \left(2\left({\tilde{\mu}\over \mu_2}
      \right)^2 +1
      +c_H^2\right)
      (C_L(z))^2
    +s_H^2  
      \left({C_L|_{z=1} 
       \over S_L|_{z=1}}\right)^2
     (S_L(z))^2 
     \right\} .
\eea
The top quark mass $m_t= k \lambda_t$  obeys
\bea
  \tilde{\mu}^2 S_R
    C_L + \mu_2^2 
     C_L \left\{
  S_R + {s_H^2\over 2S_L}\right\}
      \bigg|_{z=1, \lambda=\lambda_t} =0 .
      \label{tmass}
\eea
This equation contains  only the ratio  $\tilde{\mu}/\mu_2$
and $c$ as parameters.
As we will see below, there is
the corresponding equation for the bottom quark mass $m_b$
which contains the same parameters.
A typical set of parameters and the corresponding
quark masses are shown in Table~\ref{tab:cmu}.
The masses of quarks  in Table~\ref{tab:cmu} correspond to
the central values in the Particle Data Group review.~\cite{PDG}
\myred{(The analysis with the running masses of quarks and leptons at the
$m_Z$ scale is given in Appendix D.)}

\begin{table}[htb]
\begin{center}
\caption{The quark and lepton masses are input parameters, 
from which $c$, $|\tilde\mu/\mu_2|$ 
($|\mu_3^\ell/\tilde{\mu}^\ell|$),
and the coefficients of the wave functions are determined.
Here  $z_L=10^{15}$, $k=4.7\times 10^{17}$, 
$\theta_H ={\pi\over 2}$ and $\sin^2\theta_W=0.2312$ which
is $\overline{\textrm{MS}}$ value in the ref.~\cite{PDG}.
}
\label{tab:cmu}
\vskip 10pt
\begin{tabular}{|c|cccc|}
\hline 
& $c$ & $|\tilde{\mu}/\mu_2|$, 
\myred{$|\mu_3^\ell/\tilde{\mu}^\ell|$} & 
 \multicolumn{2}{c|}{Mass (GeV)} \\ \hline
 $W$,$Z$ & & & 80.398 & 91.1876  \\ \hline
$u$,$d$ & 0.8277 & 1.979 
   &  $2.4\times 10^{-3}$ & $4.75\times 10^{-3}$ \\
$c$,$s$ & 0.6564 & 0.08189
   & \myred{1.27} & 0.104 \\
$t$,$b$ & 0.4329 & 0.02513 
   & \myred{171.2} & \myred{4.20} \\ 
\hline 
$\nu_e$,$e$ & 0.8992 & $5.110 \times 10^{8}$ 
   &  \myred{$10^{-12}$} 
         & $0.510998910\times 10^{-3}$ \\
$\nu_\mu$,$\mu$ & 0.7354 & $1.174\times 10^{10} $
   & \myred{$9  \times 10^{-12}$} 
         & \myred{0.105658367} \\
$\nu_\tau$,$\tau$ & 0.6456 & $3.532\times 10^{10} $
   & \myred{$5.0309 \times 10^{-11}$} & 1.77684 \\ 
\hline
\end{tabular}
\vskip 7pt
\begin{tabular}{|c|ccc|}
\hline 
& $a_U$ & $a_{B+t}$ & $a_{t'}$ \\ \hline
$u$ & 
$-2.774\times 10^{-13}$ & 
$9.909 \times 10^{-14}$ & 
$5.153 \times 10^{-8}$    
\\ %
$c$ &
$-6.507\times 10^{-12}$ & 
$5.619\times 10^{-11}$  & 
$4.809\times 10^{-8}$     
\\ %
$t$ &
$-3.102\times 10^{-10}$ & 
$8.728\times 10^{-9}$   & 
$4.388\times 10^{-8}$     
\\
\hline
& $a_b$ & $a_{D+X}$ & $a_{b'}$ \\ \hline
$d$ &
$-1.401\times 10^{-13}$ & 
$1.961\times 10^{-13}$  & 
$-5.153\times 10^{-8}$     
\\ %
$s$ &
$-7.946\times 10^{-11}$ & 
$4.601\times 10^{-12}$  & 
$-4.809\times 10^{-8}$    
\\ %
$b$ &
$-1.164\times 10^{-8}$  & 
$2.069\times 10^{-10}$  & 
$-4.319\times 10^{-8}$    
\\
\hline
\noalign{\kern 7pt}
\hline 
& $a_{\nu_\tau}$ & $a_{L_{2Y}+L_{3X}}$ & $a_{\nu '_\tau}$ \\ \hline
$\nu_e$ & 
$-2.907\times 10^{-14}$ & 
$4.022\times 10^{-23}$  & 
$5.290\times 10^{-8}$     
\\ %
$\nu_{\mu}$ &
$-6.396\times 10^{-12}$ & 
$3.852\times 10^{-22}$  & 
$4.971\times 10^{-8}$     
\\ %
$\nu_{\tau}$ &
$-1.117\times 10^{-10}$ & 
$2.236\times 10^{-21}$  & 
$4.787\times 10^{-8}$     
\\
\hline
& $a_{L_{3Y}}$ & $a_{\tau + L_{1X}}$ & $a_{\tau'}$ \\ \hline
$e$ &
$-8.662\times 10^{-22}$ & 
$2.055\times 10^{-14}$  & 
$-5.290\times 10^{-8}$    
\\ %
$\mu$ &
$-3.566\times 10^{-19}$ & 
$4.523\times 10^{-12}$  & 
$-4.971\times 10^{-8}$    
\\ %
$\tau$ &
$-5.520\times 10^{-18}$ & 
$7.898\times 10^{-11}$  & 
$-4.787\times 10^{-8}$    
 \\
\hline
\end{tabular}
\end{center}
\end{table}

\subsubsection*{(ii) Bottom quark}

The bottom quark component $b(x)$ is contained in the form
\bea
   \left(\begin{array}{c}
  \tilde{b}_L \\
  {1\over \sqrt{2}} (\tilde{D}_L \pm \tilde{X}_L) \\
   \tilde{b}_L' \\
  \end{array}\right) (x,z)
   &\!\!\!=\!\!\!&
\sqrt{k} \left(\begin{array}{c}
  a_b C_L(z;\lambda, \myred{c}) \\
   a_{D\pm X} C_L(z; \lambda, \myred{c}) \\
   a_{b'} S_L(z;\lambda, \myred{c}) \\
  \end{array}\right) b_L(x)  ,
  \\
   \left(\begin{array}{c}
  \tilde{b}_R \\
  {1\over \sqrt{2}} (\tilde{D}_R \pm \tilde{X}_R) \\
   \tilde{b}_R' \\
  \end{array}\right) (x,z)
   &\!\!\!=\!\!\!&
  \sqrt{k} \left(\begin{array}{c}
  a_b S_R(z;\lambda, \myred{c}) \\
   a_{D\pm X} S_R(z; \lambda, \myred{c}) \\
   a_{b'} C_R(z;\lambda, \myred{c}) \\
  \end{array}\right) b_R(x) ~.
\eea
The $d$ and $s$ quarks are described in a similar manner.
The ratios of the coefficients are given by
\bea
 [ a_b , a_{D-X}, a_{b'} ] \,
  \myred{\simeq}  \left[
    -{\sqrt{2} \mu_2 \over \tilde{\mu}} ,
    c_H,
    {s_H C_L |_{z=1}\over S_L |_{z=1}} \right] a_{D+X} ~.
\eea
The coefficient $a_{D+X}$ is given by
\bea
  a_{D+X}^{-2}
    =\int_1^{z_L}
     dz
    \left\{
    \left(2\left({\mu_2 \over \tilde{\mu}}
      \right)^2 +1
      +c_H^2\right)
      (C_L(z))^2
    +s_H^2
      \left({C_L|_{z=1}
       \over S_L|_{z=1}}\right)^2
     (S_L(z))^2 
     \right\} .
\eea
The mass $m_b = k \lambda_b$   obeys
\bea
  \mu_2^2 S_R
    C_L + \tilde{\mu}^2 
     C_L \left\{
  S_R + {s_H^2\over 2S_L}\right\}
      \bigg|_{z=1, \lambda=\lambda_b} =0 .
      \label{bmass}
\eea

Combining Eqs.\ (\ref{tmass}) and (\ref{bmass}), one finds
\beqn
\frac{\tilde\mu^2}{\mu_2^2} 
&\!\!\!=\!\!\!&
 - \bigg\{ 1 + \frac{s_H^2}{2 S_L(1; \lambda_t, c) S_R(1; \lambda_t, c)} \bigg\} \cr
 \noalign{\kern 10pt}
&\!\!\!=\!\!\!&
 -  \bigg\{ 1 + \frac{s_H^2}{2 S_L(1; \lambda_b, c) S_R(1; \lambda_b, c)} \bigg\}^{-1}~.
   \label{muratio}
\eeqn
The value of $\theta_H$ is dynamically determined.\cite{YH1, YH2, Lim2, HHHK} 
In the present model the contributions to the effective potential $V_\eff(\theta_H)$
from multiplets containing the top quark dominate, whereas the contributions 
from other quarks and leptons are negligible. $V_\eff(\theta_H)$ has global minima
at $\theta_H= \pm \onehalf \pi$.\cite{HOOS}
Hence, given $k$ (or $z_L$) , $m_t$, $m_b$, the parameters
$c$ and $|\tilde\mu/\mu_2|$ are determined.  
These values are not 
sensitive 
on the value of $k$ very much.  See Table \ref{tab:cmu}.

The wave functions $C_L$ and $S_L$ for 
the left-handed quarks $t_L$, $b_L$
are localized near the Planck brane, whereas
$C_R$ and $S_R$ for the right-handed quarks $t_R$, $b_R$
are localized near the TeV brane.

\myred{
Generalization to the case of three generations of quarks is straightforward.
If there were no flavor mixing,  \myblue{it would be enough} 
 to replace $(\lambda_t, \lambda_b)$
in the above formulas by $(\lambda_c, \lambda_s)$ or 
$(\lambda_u, \lambda_d)$ to determine the wave functions of 
$(c,s)$ of $(u,d)$ quarks, respectively.
\myblue{The  flavor mass mixing is incorporated by considering 3-by-3 
matrices for $\mu_2$ and $\tilde \mu$, in which case  the formulas become 
more involved.}  In the present paper we evaluate electroweak gauge couplings
\myblue{with the mixing turned off}.
}

\subsubsection*{(iii) Leptons}

We need to introduce only a vector multiplet $\Psi_3$ in 
Table \ref{tab:charges} to describe a massless neutrino for each generation.
The $\tau$ lepton is  contained in the form
\bea
   \left(\begin{array}{c}
  {1\over \sqrt{2}} (\tilde{\tau}_L \pm \tilde{L}_{1XL}) \\
   \tilde{\tau}_L' \\
  \end{array}\right) (x,z)
   &\!\!\!=\!\!\!&
  \sqrt{k} \left(\begin{array}{c}
   a_{\tau \pm L_{1X}} C_L(z; \lambda, c_3) \\
   a_{\tau'} S_L(z;\lambda, c_3) \\
  \end{array}\right) \tau_L(x) ~,
  \cr
 \noalign{\kern 10pt}
   \left(\begin{array}{c}
  {1\over \sqrt{2}} (\tilde{\tau}_R \pm \tilde{L}_{1XR}) \\
   \tilde{\tau}_R' \\
  \end{array}\right) (x,z)
   &\!\!\!=\!\!\!&
  \sqrt{k} \left(\begin{array}{c}
   a_{\tau \pm L_{1X}} S_R(z; \lambda, c_3) \\
   a_{\tau'} C_R(z;\lambda, c_3) \\
  \end{array}\right) \tau_R(x) ~.
\eea
$e, \mu$ leptons are described in a similar way.
The ratios of the coefficients are given by
\bea
 [ a_{\tau-L_{1X}}, a_{\tau'} ]
  \simeq \left[
    c_H,
    {s_H C_L |_{z=1} \over S_L |_{z=1} } \right] 
 a_{\tau +L_{1X}},
\eea
for $(\mu_1^\ell)^2 \gg m_\KK$.
The coefficient $a_{\tau +L_{1X}}$ is given by
\bea
  a_{\tau +L_{1X}}^{-2}
    =\int_1^{z_L}
     dz
    \left\{
    \left(1
      +c_H^2\right)
      (C_L(z))^2
    +s_H^2
      \left({C_L |_{z=1}
       \over S_L |_{z=1} }\right)^2
     (S_L(z))^2 
     \right\} .
\eea
The mass $m_\tau = k \lambda_\tau$  is determined by
\bea
   (\mu_1^\ell )^2 
     C_L \left\{
  S_R + {s_H^2\over 2S_L}\right\}
      \bigg|_{z=1} =0 .
      \label{leptonmass1}
\eea

To describe massive neutrinos one needs to introduce two multiplets
$\Psi_3$ and $\Psi_4$.  The structure is the same as in the quark sector.
Take $c_3=c_4\equiv c$.
The left-handed  components are contained in the form
\bea
  \left(\begin{array}{c}
     \tilde{\nu}_{\tau L} \\
     {1\over \sqrt{2}}
       (\tilde{L}_{2YL} \pm \tilde{L}_{3XL}) \\
       \tilde{\nu}'_{\tau L} \\
   \end{array}\right)
   (x,z)
  &\!\!\!=\!\!\!&
  \sqrt{k} 
  \left(\begin{array}{c}
  a_{\nu_\tau} C_L (z;\lambda, c) \\
  a_{L_{2Y}\pm L_{3X}} C_L (z;\lambda, c) \\
  a_{\nu'_\tau} S_L (z;\lambda,c) \\
  \end{array}\right) 
   \nu_{\tau L} (x) ,
\nonumber
\\
   \left(\begin{array}{c}
      \tilde{L}_{3YL} \\
      {1\over \sqrt{2}}(\tilde{\tau}_L
       \pm \tilde{L}_{1XL}) \\
     \tilde{\tau}'_L \\
     \end{array}\right) 
     (x,z)
     &\!\!\!=\!\!\!&
      \sqrt{k}
      \left(\begin{array}{c}
       a_{L_{3Y}} C_L (z;\lambda, c) \\
       a_{\tau \pm L_{1X}} C_L (z;\lambda, c) \\
       a_{\tau'} S_L (z;\lambda, c) \\
     \end{array}\right)
     \tau_L(x) ~.
     \label{lepton2}
\eea
The right-handed components have the same form as in 
Eq.~(\ref{lepton2}) with
$(C_L,S_L)$ replaced by $(S_R,C_R)$. 
The equations of motion for leptons
are obtained by employing the
correspondence between leptons and quarks:
$(\nu_\tau, L_{2Y}, L_{3X}, \nu'_\tau$,
$\hat{L}_{3XR}, \hat{L}_{2YR}) 
\leftrightarrow
(U,B,t,t'$, $\hat{U}_R,\hat{B}_R)$ 
for $T^{3_L}=+1/2$,
$(L_{3YL},\tau,L_{1X},\tau'$,
$\hat{L}_{3YR},\hat{L}_{1XR})
\leftrightarrow
(b,D,X,b'$, $\hat{D}_R, \hat{X}_R)$ for $T^{3_L}=-1/2$ and
$(\mu_1^\ell, \mu_2^\ell, \mu_3^\ell ,\tilde{\mu}^\ell)
\leftrightarrow  (\mu_3, \mu_1, \tilde{\mu}, \mu_2)$.
Similarly to Eq.~(\ref{muratio}), the ratio of  the couplings is given by
\beqn
\frac{(\mu_3^\ell)^2}{(\tilde\mu^\ell)^2} 
&\!\!\!=\!\!\!&
 - \bigg\{ 1 + \frac{s_H^2}{2 S_L(1; \lambda_{\nu_\tau}, c) 
  S_R(1; \lambda_{\nu_\tau}, c)} \bigg\} \cr
 \noalign{\kern 10pt}
&\!\!\!=\!\!\!&
 -  \bigg\{ 1 + \frac{s_H^2}{2 S_L(1; \lambda_\tau, c) 
  S_R(1; \lambda_\tau, c)} \bigg\}^{-1}~.   
\eeqn 

\myred{
As in the  quark sector,  
generalization to  three generations of leptons is straightforward.
The  flavor  mixing is incorporated by taking 3-by-3 
matrices for $\mu_2^\ell$ and $\tilde \mu^\ell$.  
\myblue{The observed mixing in the neutrino sector} is large.  
In the present paper we content ourselves with considering 
the diagonal flavor.  The wave functions of the real neutrinos would 
change in the flavor space significantly, \myblue{but the effect to the gauge
couplings of charged leptons would be small provided that all neutrino 
masses are much smaller than the masses of charged leptons.}
}

Table \ref{tab:cmu} includes $c$ and $|\mu_3^\ell/\tilde{\mu}^\ell|$  for leptons
\myred{in the diagonal flavor case}. 
For neutrinos, the values $m_{\nu_e}=10^{-3}$eV and
$(m_{\nu_e},  m_{\nu_\mu},  m_{\nu_\tau})  =(m_{\nu_1},  m_{\nu_2}, m_{\nu_3})$  are taken.
The input parameters correspond to the values
derived from the mass squared differences
$\Delta m_{21}^2=8\times 10^{-5}\textrm{eV}^2$,
$\Delta m_{32}^2=2.45\times 10^{-3}\textrm{eV}^2$ which are
the central values quoted in  ref.~\cite{PDG}.
\myred{The case $\Delta m_{32}^2=- 2.45\times 10^{-3}\textrm{eV}^2$
can be analysed similarly.}
The masses of charged leptons correspond to  
the central values in ref.~\cite{PDG}.

To compare various predictions with experiments, one needs to use the 
running masses of quarks and leptons at the typical energy scale of the
problem under consideration. 
Results obtained with the input of  running masses  of quarks and leptons
at the $m_Z$ scale are summarized in Appendix~\ref{app:running}.
It is found that the coefficients of mode functions and the resulting 
gauge couplings are not very sensitive to 
the choice of these input parameters.

As in ref.~\cite{HK}, 
we define the normalized coefficients
$a_j^{\prime L,R}$ by \
\bea
  (a_U^{\prime L}, a_{B\pm t}^{\prime L}, a_{t'}^{\prime L})
    &\!\!\!=\!\!\!&
     (\sqrt{N_{C_L}}a_U,
      \sqrt{N_{C_L}}a_{B\pm t},
      \sqrt{N_{S_L}}a_{t'}) ,
\nonumber
\\
  (a_U^{\prime R}, a_{B\pm t}^{\prime R}, a_{t'}^{\prime R})
    &\!\!\!=\!\!\!&
     (\sqrt{N_{S_R}}a_U,
      \sqrt{N_{S_R}}a_{B\pm t},
      \sqrt{N_{C_R}}a_{t'}) ,
\eea
where $N_{C_L}=\int_1^{z_L} dz C_L (z;\lambda,c)^2$, etc.
These coefficients satisfy 
$|a^{\prime L}_U|^2 + |a^{\prime L}_{B+t}|^2 + |a^{\prime L}_{B-t}|^2 + |a^{\prime L}_{t'}|^2 =1$
etc. \myred{so that} 
$|a^{\prime L, R}_j|^2$ represents the weight of the $j$-th component in each quark or lepton.
They are tabulated in Table \ref{tab:aprimequark} and Table \ref{tab:aprimelepton}.
At $\theta_H= \onehalf \pi$, all of the $a^{\prime}_{B-t}$, $a^{\prime}_{D-X}$,
$a^{\prime}_{L_{1Y}-L_{2X}}$, and $a^{\prime}_{\tau-L_{1X}}$ components vanish.

\begin{table}[htb]
\begin{center}
\caption{Normalized coefficients in mode expansion 
for quarks.}
\label{tab:aprimequark}
\vskip 5pt
\begin{tabular}{|c|ccc|ccc|}
\hline 
& $a'_{UL}$ & $a'_{B+t \, L}$ & $a'_{t'L}$ 
& $a'_{UR}$ & $a'_{B+t \, R}$ & $a'_{t'R}$
\\ \hline
$u$ & 
$-0.8925$               & 
0.3189                  & 
0.3189                  & 
$-2.370\times 10^{-11}$ & 
$8.468\times 10^{-12}$  & 
1 
\\ 
$c$ &
$-0.08162$             & 
0.7047                 & 
0.7047                 & 
$-2.815\times 10^{-7}$ & 
$2.430\times 10^{-6}$  & 
1
\\
$t$ &
$-0.02582$             & 
0.7265                 & 
0.6867                 & 
$-0.001769$            & 
0.04976                & 
0.9988                   
\\
\hline
\noalign{\kern 3pt}
\hline
& $a'_{bL}$ & $a'_{D+XL}$ & $a'_{b'L}$
& $a'_{bR}$ & $a'_{D+XR}$ & $a'_{b'R}$ 
\\ \hline
$d$ & 
$-0.4510$               & 
0.6311                  & 
$-0.6311$               & 
$-2.370\times 10^{-11}$ & 
$3.317\times 10^{-11}$  & 
$-1$
\\ 
$s$ &
$-0.9967$              & 
0.05771                &  
$-0.05771$             & 
$-2.815\times 10^{-7}$ & 
$1.630\times 10^{-8}$  & 
$-1$
\\ 
$b$ &
$-0.9997$   & 
0.01777     & 
$-0.01685$  & 
$-0.001648$ & 
0.00002930  & 
$-1$            
\\
\hline
\end{tabular}
\end{center}
\end{table}

\begin{table}[htb]
\begin{center}
\caption{Normalized coefficients in mode expansion 
for leptons.}
\label{tab:aprimelepton}
\vskip 5pt
\begin{tabular}{|c|ccc|ccc|}
\hline  
& $a'_{\nu_\tau L}$ & $a'_{L_{2Y}+L_{3X} L}$ & $a'_{\nu '_\tau L}$ 
& $a'_{\nu_\tau R}$ & $a'_{L_{2Y}+L_{3X} R}$ & $a'_{\nu '_\tau R}$ 
\\ \hline
$\nu_e$ & 
$-1$ &
$1.384\times\!\! 10^{-9}$   & 
$1.384\times\!\! 10^{-9}$   & 
$-1.066\times\!\! 10^{-21}$ & 
$1.476\times\!\! 10^{-30}$  & 
1
\\ 
$\nu_{\mu}$ &
$-1$ &
$6.023\times\!\! 10^{-11}$  & 
$6.023\times\!\! 10^{-11}$  & 
$-1.993\times\!\! 10^{-18}$ & 
$1.200\times\!\! 10^{-28}$  & 
1
\\ 
$\nu_{\tau}$ &
$-1$ &
$2.002\times\!\! 10^{-11}$  & 
$2.002\times\!\! 10^{-11}$  & 
$-1.911\times\!\! 10^{-16}$ & 
$3.825\times\!\! 10^{-27}$  & 
1
\\
\hline
\noalign{\kern 3pt}
\hline
& $a'_{L_{3Y} L}$ & $a'_{\tau + L_{1X} L}$ & $a'_{\tau' L}$
& $a'_{L_{3Y} R}$ & $a'_{\tau +L_{1X} R}$ & $a'_{\tau' R}$ 
\\ \hline
$e$ &
$-2.980\times\!\! 10^{-8}$  & 
0.7071                      & 
$-0.7071$                   & 
$-1.624\times\!\! 10^{-20}$ & 
$3.853\times\!\! 10^{-13}$  & 
$-1$
\\ 
$\mu$ & 
$-5.576\times\!\! 10^{-8}$  & 
0.7071                      & 
$-0.7071$                   & 
$-1.304\times\!\! 10^{-15}$ & 
$1.654\times\!\! 10^{-8}$   & 
$-1$
\\ 
$\tau$ &
$-4.942\times\!\! 10^{-8}$  & 
0.7071                      & 
$-0.7071$                   & 
$-3.335\times\!\! 10^{-13}$ & 
$4.772\times\!\! 10^{-6}$   & 
$-1$
\\
\hline
\end{tabular}
\end{center}
\end{table}
%

The left-handed quarks have contributions from
both  singlet and non-singlet components of $SU(2)_L\times SU(2)_R$.
The left-handed neutrinos have contributions almost entirely 
from the $T^{3_L}=T^{3_R}=+1/2$ component.        
The left-handed charged leptons receive contributions from
the $T^{3_L}=-T^{3_R}=\pm 1/2$ component   and the  singlet component.      
The right-handed quarks and leptons are
dominated by the singlet component.

Regarding the isospin eigenvalues in $SU(2)_L\times SU(2)_R$,
all the nonvanishing components are found to be left-right symmetric.
Indeed the eigenvalues are given by
$T^{3_L}=T^{3_R}=+1/2$ for $(U,\nu_\tau)$,
$T^{3_L}=T^{3_R}=-1/2$ for $(b,L_{3Y})$,
$T^{3_L}=T^{3_R}=0$ for
$(t', b', \nu'_\tau, \tau')$
and
$(T^{3_L}=-T^{3_R}=+1/2)\oplus
(T^{3_L}=-T^{3_R}=-1/2)$ for
$(B+t, D+X, L_{2Y}+L_{3X}, \tau+L_{1X})$.
The left-right symmetry is preserved 
as the left-right antisymmetric part vanishes at $\theta_H=\pm \pi/2$.
As expected from the situation in the left-right symmetric models,
we will see in the next section that the gauge couplings
deviate little   from those in the standard model.

\section{Electroweak currents}

Inserting the wave functions of the 4D gauge fields $A_\mu^\gamma(x)$,
$W_\mu(x)$ and $Z_\mu(x)$ into the 5D gauge couplings,  one finds that
\bea
 &&\hskip -1cm   
 g_A \tilde{A}_\mu(x,z) 
 +g_B Q_X \tilde{B}_\mu (x,z)
 =  e A_\mu^\gamma(x)  Q_E \cr
\noalign{\kern 10pt}
&& \hskip .5cm
+ g_A W_\mu(x) \bigg\{ N_W(z)   T^{-} (\theta_H)
 - D_W (z) \frac{\sin\theta_H}{\sqrt{2}} T^{\hat{-}} \bigg\} \cr
\noalign{\kern 10pt}
&& \hskip .5cm
+ g_A W_\mu^\dagger (x) \bigg\{ N_W(z)   T^{+}(\theta_H)
 - D_W (z) \frac{\sin\theta_H}{\sqrt{2}} T^{\hat{+}} \bigg\} \cr
\noalign{\kern 10pt}
&& \hskip .5cm
+ \frac{g_A Z_\mu (x)}{\cos \theta_W}
 \bigg\{ N_Z(z) \Big( T^{3}(\theta_H) - \sin^2 \theta_W \, Q_E\Big) 
 - D_Z (z) \frac{\sin\theta_H}{\sqrt{2}} T^{\hat{3}} \bigg\} ~, \cr
\noalign{\kern 10pt}
&& \hskip -1.cm
T^a(\theta_H) = \frac{1+ \cos\theta_H}{2}  T^{a_L}
+ \frac{1- \cos\theta_H}{2}  T^{a_R} ~,~~
T^\pm = \frac{T^{1} \pm i T^{2}}{\sqrt{2}} ~.
\label{current1}
\eea
The wave functions $N_{W, Z}(z)$ and $D_{W, Z}(z)$ have been defined in the previous
section.  Explicit forms of gauge couplings with fermions in the vector representation of 
$SO(5)$  are given in Appendix~\ref{ap:fermi}.
The observed weak $SU(2)$ is the mixture of the original $SU(2)_L$ and 
$SU(2)_R$.  There appear $T^{\hat a}$ components as well.

4D gauge couplings of quarks and leptons are obtained by inserting their
wave functions into $\tilde \Psi_a (x,z)$ and integrating 
\[
\sum_a \int_1^{z_L} \frac{dz}{k} \, \overline{\tilde \Psi}_a \Gamma^\mu 
(g_A  \tilde{A}_\mu + g_B Q_{X a} \tilde{B}_\mu) \tilde \Psi_a ~.
\]
in (\ref{fermiact}).
As they appear as overlap integrals, the 4D gauge couplings are not
universal and are expected to deviate from those in the standard model
at nonvanishing $\theta_H$.  
Surprisingly the deviation turns out very small, 
although the wave functions have significant dependence on $\theta_H$.
Useful formulas for evaluating 4D gauge-fermion couplings are given in 
Appendix \ref{ap:gauge-couple}.

\ignore{We analyze explicitly couplings of photon, $W$ and $Z$ bosons with 
quarks and leptons.}

\subsubsection*{(i) Photon couplings}

The photon couplings are universal.  The $U(1)_\EM$ invariance remains intact.
The wave function of the photon is constant with respect to 
$z$ so that the $z$-integrals for the gauge couplings 
are fixed by the normalization of the fermion wave functions.  
For the third generation one finds that
\bea
  && eA_\mu^\gamma
 \bigg\{ {2\over 3}(\bar{t}_L\gamma^\mu t_L
  +\bar{t}_R \gamma^\mu t_R)
 -{1\over 3}(\bar{b}_L \gamma^\mu b_L
  +\bar{b}_R \gamma^\mu b_R) 
 - (\bar{\tau}_L \gamma^\mu  \tau_L
 + \bar{\tau}_R \gamma^\mu \tau_R)  \bigg\} ~ .
 \label{photoncoupling1}
\eea

\subsubsection*{(ii) $W$ boson couplings}

There arise deviations in the $W$ boson couplings from the standard model.
For the $t$ and $b$ quarks their interactions are given by 
\bea
{1\over 2} g^{(W)}_{tb,L} 
  (W_\mu  \bar{b}_L \gamma^\mu t_L 
  +W_\mu^\dag  \bar{t}_L \gamma^\mu b_L) 
 +  {1\over 2} g^{(W)}_{tb, R}
  (W_\mu  \bar{b}_R \gamma^\mu t_R 
  +W_\mu^\dag  \bar{t}_R \gamma^\mu b_R).
\label{Wcoupling1}
\eea
Not only left-handed components but also right-handed  components
couple to $W$ in general.  The couplings are given by
\bea   
g^{(W)}_{tb,L}  &\!\!\!=\!\!\!&
  g_A\int_1^{z_L} dz   \,  \bigg[
   N_W C_L(\lambda_b) C_L(\lambda_t)
  \Big\{  (a_{D+X} +a_{D-X})  a_U  + a_b (a_{B+t}- a_{B-t}) \Big\}  \cr
\noalign{\kern 10pt}
&& 
- (1-\cos\theta_H)  N_W C_L(\lambda_b) C_L (\lambda_t) 
     (a_{D-X}  a_U  -a_b  a_{B-t}) \cr
\noalign{\kern 10pt}
&&  - \sin\theta_H  D_W 
\Big\{ C_L(\lambda_b)  S_L(\lambda_t) a_b a_{t'}  -S_L(\lambda_b)
         C_L (\lambda_t) a_{b'} a_U \Big\}  \bigg]  ~.
\label{Wcoupling2}
\eea
Here $N_W=N_W(z)$, $C_L(\lambda)= C_L(z; \lambda)$ etc.  
The formula for $g^{(W)}_{tb,R}$ is obtained from (\ref{Wcoupling2}) by
replacing $C_L(\lambda)$ and $S_L(\lambda)$ by $S_R(\lambda)$ and $C_R(\lambda)$,
respectively.

We have tabulated the numerical values of these couplings at $\theta_H= \onehalf \pi$
in Table~\ref{tab:gw}.  In the table the normalized couplings
\beeq
g'^{(W)}_{f, LR} = \frac{1}{\myred{g_W}} g^{(W)}_{f, LR} ~~, ~~
\myred{g_W} =  g^{(W)}_{e \nu, L} 
\label{Wcoupling3}
\eneq
are also listed where \myred{$g_W$}
is the measured $W e \nu_e$ coupling, corresponding to 
the $SU(2)_L$ coupling in the standard model.
The origin of the value \myred{$g^{(W)}_{e \nu, L} / \bar g_A= 1.00533$, 
where $\bar g_A$ is the 4D gauge coupling defined in (\ref{4Dcoupling1}),}  
is traced back to $\sqrt{L}N_W \simeq 1.00533$ as described in Section 4.1.

\begin{table}[htb]
\begin{center}
\caption{The $W$ couplings $g^{(W)}_{f, LR}$ and $g'^{(W)}_{f, LR} =g^{(W)}_{f, LR}/\myred{g_W} $ 
at $\theta_H=\onehalf \pi$. 
($\myred{g_W} \equiv g^{(W)}_{e \nu L}$.)  
Here the parameters given in Table~\ref{tab:cmu} 
are used.}
\label{tab:gw}
\vskip 10pt
\begin{tabular}{|c|ccc|}  
\hline 
      $f$      & $u,d$   & $c,s$  & $t,b$ \\ \hline
$g^{(W)}_{fL}/\myred{\bar{g}_A}$ 
     & 1.00533 & 1.00533 & 0.98192 \\ 
$g^{(W)}_{fR}/\myred{\bar{g}_A}$ 
    & $-1.399\times\!\! 10^{-11}$ & 
$-1.906\times\!\! 10^{-7}$  & 
$-0.001271$                   
 \\ 
  \hline
  $g'^{(W)}_{fL}$ & 1. & 1. & 0.9767 \\ 
$g'^{(W)}_{fR}$ & $-1.392\times\!\! 10^{-11}$ & 
$-1.896\times\!\! 10^{-7}$  & 
$-0.001264$                   
\\ 
\hline
\noalign{\kern 7pt}
\hline
       $f$     & $\nu_e,e$   & $\nu_{\mu},\mu$  & $\nu_{\tau},\tau$ \\ \hline
$g^{(W)}_{fL}/\myred{\bar{g}_A}$ 
    & 1.00533 & 1.00533 & 1.00533 \\ 
$g^{(W)}_{fR}/\myred{\bar{g}_A}$ 
 & $-4.787\times\!\! 10^{-21}$ & 
$-4.163\times\!\! 10^{-16}$ & 
$-1.139\times\!\! 10^{-13}$   
 \\ 
  \hline
  $g'^{(W)}_{fL}$ & 1. & 1. & 1. \\ 
\myred{$g'^{(W)}_{fR}$}
   & $-4.761\times\!\! 10^{-21}$ & 
$-4.141\times\!\! 10^{-16}$ & 
$-1.133\times\!\! 10^{-13}$   
\\ 
\hline
\end{tabular}
\end{center}
\end{table}
\begin{table}[htb]
\begin{center}
\caption{The deviation $g'^{(W)}_{fL}-1$ of the $W$ couplings
from the standard model, which represents the violation of the universality. }
\label{tab:gw2}
\vskip 8pt
\begin{tabular}{|c|ccccc|}
\hline
&\multicolumn{5}{c|}{$g'^{(W)}_{fL}-1$} \\ 
\hline 
 $f$   & $\nu_{\mu},\mu$  & $\nu_{\tau},\tau$ & $u,d$ & $c,s$  & $t,b$  \\ \hline
&
$-1.022\times\!\! 10^{-8}$  & 
$-2.736\times\!\! 10^{-6}$   & 
$-2.778\times\!\! 10^{-11}$ & 
$-1.415\times\!\! 10^{-6}$  & 
$-0.02329$ 
 \\ \hline
\end{tabular}
\end{center}
\end{table}

Notice that the couplings of left-handed quarks and leptons are close to \myred{$g_W$}
except for the top quark.  To see the tiny violation of the weak universality
we have listed $g'^{(W)}_{fL}-1$ in Table \ref{tab:gw2}.
\myred{By definition, $g'^{(W)}_{\nu_e, e \,L}-1=0$.}
The violation of the $\mu$-$e$ universality is of $-1.022 \times 10^{-8}$
\myred{for $z_L = 10^{15}$},
which is below the current limit.
Similar behavior has been previously reported in the $SU(3)$ model in ref.\ \cite{HNSS}.
The violation becomes appreciable for the $tb$ coupling.
The couplings of right-handed quarks and leptons are nonvanishing,
but are all tiny.

\subsubsection*{(iii) $Z$ boson couplings}

The $Z$ boson couplings of $t$ and $b$ quarks take the form
\beqn
&&\hskip -1cm
\frac{1}{\cos\theta_W} Z_\mu \bigg\{ 
g^{(Z)}_{tL}   \bar{t}_L \gamma^\mu t_L
+g^{(Z)}_{tR} \bar{t}_R \gamma^\mu t_R  
+g^{(Z)}_{bL} \bar{b}_L \gamma^\mu b_L
+g^{(Z)}_{bR} \bar{b}_R \gamma^\mu b_R   \bigg\}, \cr
\noalign{\kern 10pt}
&&\hskip 1.cm
g^{(Z)}_{tL,R}= + \frac{1}{2}g^{(Z)3}_{tL,R}
- \frac{2}{3} g^{(Z)Q}_{tL, R} \sin^2 \theta_W ~, \cr
\noalign{\kern 10pt}
&&\hskip 1.cm
g^{(Z)}_{bL,R}= - \frac{1}{2}g^{(Z)3}_{bL,R}
+ \frac{1}{3} g^{(Z)Q}_{bL, R} \sin^2 \theta_W ~.
\label{Zcoupling1}
\eeqn
Here the couplings $g^{(Z)3}$ and $g^{(Z)Q}$ are given by
\bea
g_{tL}^{(Z)3}  &\!\!\!=\!\!\!&g_A \int_1^{z_L} dz \,
 \Big\{  N_Z (C_L(\lambda_t))^2 
 \big(   a_U^2  -2\cos\theta_H a_{B+t} a_{B-t}  \big)  \cr
\noalign{\kern 5pt}
&& \hskip 3.5cm  
-2\sin\theta_H D_Z C_L(\lambda_t) S_L(\lambda_t)  a_{B+t} a_{t'} \Big\} ,  \cr
\noalign{\kern 10pt}
g_{tL}^{(Z)Q} &\!\!\!=\!\!\!&
 g_A \int_1^{z_L} dz \,  N_Z  \Big\{ (C_L(\lambda_t))^2
   \big( a_U^2 +a_{B+t}^2 +a_{B-t}^2 \big)
   +(S_L(\lambda_t))^2  a_{t'}^2 \Big\} , \cr
\noalign{\kern 10pt}
g_{bL}^{(Z)3}    &\!\!\!=\!\!\!&g_A \int_1^{z_L} dz \,
\Big\{  N_Z (C_L(\lambda_b))^2 
\big( a_b^2  +2\cos\theta_H a_{D+X} a_{D-X}  \big) \cr
\noalign{\kern 5pt}
&& \hskip 3.5cm  
 +2\sin\theta_H D_Z C_L(\lambda_b) S_L(\lambda_b)  a_{D+X} a_{b'} \Big\} , \cr
\noalign{\kern 10pt}
g_{bL}^{(Z)Q}  &\!\!\!=\!\!\!&  g_A \int_1^{z_L} dz \,  N_Z 
  \Big\{(C_L(\lambda_b))^2 
  \big( a_b^2 +a_{D+X}^2 +a_{D-X}^2 \big) +(S_L(\lambda_b))^2  a_{b'}^2 \Big\} .
\label{Zcoupling2}   
\eea
The formulas for $g_{tb, R}^{(Z)3}$ and $g_{tb, R}^{(Z)Q}$ are 
obtained from (\ref{Zcoupling2}) by
replacing $C_L(\lambda)$ and $S_L(\lambda)$ by $S_R(\lambda)$ and $C_R(\lambda)$,
respectively.

%
%
%
\begin{table}[htb]
\begin{center}
\caption{$Z$ boson couplings of quarks  
at $\theta_H = \onehalf \pi$.
For entry denoted as SM, 
$g^{(Z)}_{tL}/\myred{\bar{g}_A}$ and 
$g^{(Z)}_{bL}/\myred{\bar{g}_A}$ are 
${1\over 2}-{2\over 3}\sin^2\theta_W$ and
$-{1\over 2}+{1\over 3}\sin^2\theta_W$,
respectively. } 
\label{tab:gz-q}
\vskip 10pt
\begin{tabular}{|c|ccc|c|} 
\hline 
 $f$ & $u$ & $c$ & $t$  & SM \\ \hline
$g^{(Z)}_{fL}/\myred{\bar{g}_A}$  
& 0.3483 & 
0.3483 & 
0.3221   
 & 0.3459 \\
$g^{(Z)}_{fR}/\myred{\bar{g}_A}$ 
& $-0.1564$ & 
$-0.1564$ & 
$-0.1833$   
 & $-0.1541$ \\
$g^{(Z)3}_{fL}/\myred{\bar{g}_A}$
 & 1.00699 & 
1.00698 & 
0.9549    
 & 1
 \\
$g^{(Z)3}_{fR}/\myred{\bar{g}_A}$
& $-7.086\times\!\! 10^{-12}$ & 
$-2.332\times\!\! 10^{-6}$  & 
$-0.05400$              
 & 0
\\
$g^{(Z)Q}_{fL}/\myred{\bar{g}_A}$ 
 &1.00699 & 
1.00699 & 
1.0079    
 & 1
\\
$g^{(Z)Q}_{fR}/\myred{\bar{g}_A}$
 & 1.0141 & 
1.0145 & 
1.0141   
 &1
 \\
\hline
\noalign{\kern 5pt}
\hline
 $f$ &  $d$ & $s$ & $b$  & SM \\ \hline
$g^{(Z)}_{fL}/\myred{\bar{g}_A}$  
  & $-0.4259$ & 
$-0.4259$ & 
$-0.4264$   
 & $-0.4229$ \\
$g^{(Z)}_{fR}/\myred{\bar{g}_A}$ 
 & 0.07820 & 
0.07818&  
0.07816   
  & 0.07707 \\
$g^{(Z)3}_{fL}/\myred{\bar{g}_A}$
 & 1.00699 & 
1.00699 & 
1.00813   
 & 1\\
$g^{(Z)3}_{fR}/\myred{\bar{g}_A}$
 & $-2.776\times\!\! 10^{-11}$ & 
$-1.564\times\!\! 10^{-8}$  & 
$-0.00003003$                 
 & 0
\\
$g^{(Z)Q}_{fL}/\myred{\bar{g}_A}$ 
  & 1.00699 & 
1.00699 & 
1.00816   
 & 1
\\
$g^{(Z)Q}_{fR}/\myred{\bar{g}_A}$
  & 1.0148 & 
1.0145 & 
1.0140   
 & 1 \\
\hline 
\end{tabular}
\end{center}
\end{table}
\vskip 5pt
\begin{table}[htb]
\begin{center}
\caption{$Z$ boson couplings of leptons
at $\theta_H = \onehalf \pi$.
For entry denoted as SM, 
$g^{(Z)}_{\nu_\tau L}/\myred{\bar{g}_A}$ and 
$g^{(Z)}_{\tau L}/\myred{\bar{g}_A}$ are 
${1\over 2}$ and $-{1\over 2}+\sin^2\theta_W$,
respectively. } 
\label{tab:gz-l}
\vskip 10pt
\begin{tabular}{|c|ccc|c|} 
\hline 
 $f$ & $\nu_e$ & $\nu_\mu$ & $\nu_\tau$  & SM \\ \hline
$g^{(Z)}_{fL}/\myred{\bar{g}_A}$  & 
0.5035 & 
0.5035 & 
0.5035   
 & 0.5 \\
$g^{(Z)}_{fR}/\myred{\bar{g}_A}$ & 
$-5.785\times\!\! 10^{-31}$ & 
$-5.421\times\!\! 10^{-29}$ & 
$-1.850\times\!\! 10^{-27}$   
 & $0$ \\
$g^{(Z)3}_{fL}/\myred{\bar{g}_A}$ & 
1.00699 & 
1.00699 & 
1.00699   
 & 1
 \\
$g^{(Z)3}_{fR}/\myred{\bar{g}_A}$ & 
$-1.157\times\!\! 10^{-30}$ & 
$-1.084\times\!\! 10^{-28}$ & 
$-3.700\times\!\! 10^{-27}$   
 & 0
\\
$g^{(Z)Q}_{fL}/\myred{\bar{g}_A}$  &
1.00699 & 
1.00699 & 
1.00699   
 & 1
\\
$g^{(Z)Q}_{fR}/\myred{\bar{g}_A}$ &
1.0149 & 
1.0146 & 
1.0145  
 &1
 \\
\hline 
\noalign{\kern 5pt} 
\hline
 $f$ &  $e$ & $\mu$ & $\tau$  & SM \\ \hline
$g^{(Z)}_{fL}/\myred{\bar{g}_A}$  & 
$-0.2707$ & 
$-0.2707$ & 
$-0.2707$   
 & $-0.2688$ \\
$g^{(Z)}_{fR}/\myred{\bar{g}_A}$  & 
$0.2346$ & 
$0.2346$ & 
$0.2345$   
  & 0.2312 \\
$g^{(Z)3}_{fL}/\myred{\bar{g}_A}$ & 
1.00699 & 
1.00699 & 
1.00698   
 & 1\\
$g^{(Z)3}_{fR}/\myred{\bar{g}_A}$ & 
$-3.021\times\!\! 10^{-13}$ & 
$-1.494\times\!\! 10^{-8}$  & 
$-4.615\times\!\! 10^{-6}$    
 & 0
\\
$g^{(Z)Q}_{fL}/\myred{\bar{g}_A}$   & 
1.00699 & 
1.00699 & 
1.00699   
 & 1
\\
$g^{(Z)Q}_{fR}/\myred{\bar{g}_A}$  & 
1.0149 & 
1.0146 & 
1.0145   
 & 1 \\
 \hline 
\end{tabular}
\end{center}
\end{table}

The values of $g^{(Z)3}/\myred{\bar{g}_A}$ and 
$g^{(Z)Q}/\myred{\bar{g}_A}$ for each quark 
and lepton are listed in 
Tables~\ref{tab:gz-q} and \ref{tab:gz-l}, respectively.
The origin of the value $g_{uL}^{(Z)3}/\myred{\bar{g}_A} = 1.00699$
is traced back to the value 
$\sqrt{L}N_Z \simeq 1.00699$.  
As emphasized in the discussion of the $W$ coupling, the observed weak coupling is
\myred{$g_W$} defined in (\ref{Wcoupling3}).  
Accordingly we define
\beeq
g^{\prime (Z)}_{f, LR} = \frac{1}{\myred{g_W}} 
\,  g^{(Z)}_{f, LR}  ~.
\label{Zcoupling3}
\eneq
The numerical values are listed in Table \ref{tab:gz2}.  
To emphasize the magnitude of deviation the quantity
\beeq
\frac{g^{\prime (Z)}_{f, LR}}{g^{(Z)}_{SM}} - 1
\label{Zcoupling4}
\eneq
is also given in  parentheses in the table, where $g^{(Z)}_{SM}$ is the coupling in the 
standard model.

\begin{table}[htb]
\begin{center}
\caption{$Z$ boson couplings $g'^{(Z)}_{fL,R}$ in (\ref{Zcoupling3}) 
and the deviation from SM.  The quantity (\ref{Zcoupling4}), which measures the 
deviation from the standard model,  is given in parentheses.}
\label{tab:gz2}
\vskip 10pt
\begin{tabular}{|c|ccc|c|} 
\hline 
 $f$ & $u$ & $c$ & $t$ & SM  \\ \hline
$g'^{(Z)}_{fL}$  & 
0.3464 & 
0.3464 & 
0.3204 & 
0.3459   
\\
& 
(+0.001644) & 
(+0.001640) & 
$(-0.07361)$ & 
 \\
$g'^{(Z)}_{fR}$
 & $-0.1556$ & 
$-0.1556$ & 
$-0.1823$ & 
$-0.1541$   
\\
& 
(+0.009382) & 
(+0.009100) & 
(+0.1829)   & 
\\ \hline
\noalign{\kern 5pt}
\hline
$f$  & $d$ & $s$ & $b$ & SM  \\ \hline
$g'^{(Z)}_{fL}$  & 
$-0.4236$ & 
$-0.4236$ & 
$-0.4241$ & 
$-0.4229$   
\\
& 
(+0.001644) & 
(+0.001644) & 
(+0.002775) & 
 \\
$g'^{(Z)}_{fR}$ & 
0.07779 & 
0.07777 & 
0.07774 & 
0.07707   
\\
& 
(+0.009382) & 
(+0.009092) & 
(+0.008786) & 
\\ \hline
\noalign{\kern 5pt}
\hline 
 $f$ & $\nu_e$ & $\nu_{\mu}$ & $\nu_{\tau}$ & SM  \\ \hline
$g'^{(Z)}_{fL}$  & 
0.500822 & 
0.500822 & 
0.500822 & 
0.5
\\
& 
(+0.001644) & 
(+0.001644) & 
(+0.001644) & 
 \\
$g'^{(Z)}_{fR}$
 & 
 $-5.754\times\!\! 10^{-31}$ & 
$-5.392\times\!\! 10^{-29}$ & 
$-1.840\times\!\! 10^{-27}$ & 
 0
 \\
          & $(-)$
 & $(-)$
 & $(-)$ &
\\ \hline
\noalign{\kern 5pt}
\hline 
 $f$ & $e$ & $\mu$ & $\tau$ & SM  \\ \hline
$g'^{(Z)}_{fL}$  & 
$-0.2692$ & 
$-0.2692$ & 
$-0.2692$ & 
$-0.2688$   
\\
& 
(+0.001644) & 
(+0.001644) & 
(+0.001633) & 
 \\
$g'^{(Z)}_{fR}$
 & 
0.2334 & 
0.2333 & 
0.2333 & 
0.2312   
\\
& 
(+0.009485) & 
(+0.009234) & 
(+0.009082) & 
\\ \hline
\end{tabular}
\end{center}
\end{table} 

One can see from the tables, the deviation from the standard model is rather small
(0.1\% to 1\%)  except for the top quark coupling \myred{for $z_L = 10^{15}$}.  
The $Z$ couplings of  the left- and right-handed top quark deviate from those in 
the standard model by \myred{$-7$\%} and 18\%, respectively.
The deviations in the $Z  b_L \bar b_L $ and $Z b_R \bar b_R$ couplings 
are 0.3\% and 0.9\%, respectively.
The universality in the $Z$ couplings is slightly violated as in the $W$ couplings.

\section{Discussions and conclusion}

We have determined the $W$ and  $Z$  
currents in the $SO(5)\times U(1)_X$ model with
three generations of quarks and leptons \myred{with  no flavor mixing},
which generalizes the model of ref.~\cite{HOOS}. 
The electroweak currents depend on profiles
of the wave functions of $W$, $Z$, quarks and leptons
in the fifth dimension and in the $SO(5)$ group.
Despite their highly nontrivial profiles,
it has been found that the deviation of the couplings of
quarks and leptons to gauge bosons from those in the standard model
are less than $1 \%$ except for the top quark.
The largest deviation appears in 
the $Z$ boson coupling of the  right-handed top quark, 
amounting to $\sim 18\%$.
The next largest deviation is in
the $Z$ boson coupling of the  left-handed top quark, \myred{$\sim -7\%$}.
The $W$ boson coupling of $t$- and $b$-quarks
has deviation of $\sim 2\%$.
In addition, the universality of  the gauge couplings in the flavor is violated. 
The violation of the universality for the first two generations
is very small, \myred{suppressed by many orders of magnitude, and is consistent with 
the current experimental bounds.}
The largest violation of the universality arises in
the $Z$ boson coupling for the right-handed top quark \myred{to be}  $\sim 18\%$. 

The reason for the small deviation from the standard model 
lies in the  left-right symmetry as discussed
in a general analysis in ref.~\cite{Agashe:2006at}.
The normalized coefficients,  $a'$, of mode functions 
are nonvanishing only for the part symmetric  under the exchange of 
the left- and right-isospin eigenvalues in $SU(2)_L\times SU(2)_R$.
The part antisymmetric  under the left-right exchange
has a coefficient  proportional to $\cos\theta_H$,  
which \myred{exactly} vanishes at $\theta_H=\pm {1\over 2}\pi$. 
The effective potential $V_\eff (\theta_H)$ is minimized at 
$\theta_H=\pm {1\over 2}\pi$ 
in the $SO(5)\times U(1)_X$ model in warped space 
\myred{provided multiplets in the bulk containing 
quarks and leptons} belong to the  vectorial representations of  $SO(5)$.\cite{HKT}

In investigating the  gauge couplings of quarks and leptons,
we have not considered, in the present paper,  the flavor mixing in
quarks and leptons.
The flavor mixing can be  incorporated by replacing
the couplings $\mu_\alpha$, $\mu_\alpha^\ell$,
$\tilde{\mu}$, $\tilde{\mu}^\ell$
in the brane coupling  (\ref{brane1})   by general 
matrices with flavor indices~\cite{Nomura1,Uekusa:2008iz}.
\myred{
In the neutrino sector, in particular, the mixing is large and important.
Also in the quark sector the CKM mixing must be taken into account to
accurately evaluate the amount of the violation of the weak universality.
}
\myblue{
Further  the matrices
$\mu_\alpha$, $\mu_\alpha^\ell$, $\tilde{\mu}$, $\tilde{\mu}^\ell$
need to be constrained, for the consistency with the absence of 
 FCNC (flavor-changing  neutral currents), 
for instance,  via a version of the Glashow-Iliopoulos-Maiani mechanism.\cite{%
Agashe:2004cp, Fitzpatrick:2007sa, Csaki:2008zd}
Elaboration of the analysis in this direction 
is urgent.
}

Further,  in the electroweak precision tests,
gauge couplings of fermions have been estimated to extreme accuracy
with respect to, for instance,   the forward-backward production asymmetry
on the $Z$ resonance.
As the deviation of  the couplings  from the standard model is very
small except for the top quark,
the deviation for the asymmetry production are 
also expected to be small except for the top quark.
\myred{More detailed analysis on this matter is deferred to} a separate paper~\cite{ue}.

\myblue{
QCD gauge interactions are implicit in our model, and 
color $SU(3)$ should be included 
as a direct product group in the bulk symmetry.
KK gluons in color $SU(3)$ would give rise to important  QCD corrections 
to the electroweak gauge couplings.~\cite{Lillie:2007yh, Cho:2009rj}
These effects need to be taken into account appropriately. 
}

In conclusion, we have presented  the $SO(5) \times U(1)$ gauge-Higgs 
unification model in the RS space which yields the electroweak gauge couplings 
of quarks and leptons very close to the observed couplings 
\myred{and consistent with  the current data.}
There appear appreciable violation of the universality in the top and bottom quark
couplings, \myred{which awaits  confirmation in collider experiments in the future}.
The scenario \myred{leads to} a new view for the Higgs boson.  
It is a part of the gauge fields in the extra dimension.
\myred{While it gives masses to quarks and leptons, it becomes absolutely stable.}
Higgs bosons can be the cold dark matter in the universe.\cite{HKT}
The model has to be examined in more  detail to explore  experimental 
and observational consequences 
\myred{
by taking  account of the flavor mixing and radiative corrections}.

\vspace{3ex}

\subsubsection*{Acknowledgments}
We would like to thank Y. Koide and M. Tanaka for many valuable discussions.
This work was supported in part 
by  Scientific Grants from the Ministry of Education and Science, 
Grant No.\ 20244028, Grant No.\ 20025004,  and Grant No.\ 50324744.

\vskip 1cm

\begin{appendix}

\section{Boundary conditions for gauge bosons
at $z=1$ \label{ap:bcg}}
In the twisted gauge the boundary conditions for gauge bosons at $z=1$
are given by
\bea
   0&\!\!\!=\!\!\!&
      -\sin\theta_H \,\tilde{A}_\mu^{a_L}
      +\sin\theta_H \,\tilde{A}_\mu^{a_R}
      -\sqrt{2}\cos\theta_H \,\tilde{A}_\mu^{\hat{a}} ~,
\cr
  0&\!\!\!=\!\!\!&
    \cos^2{\theta_H\over 2}\,
     \partial_z \tilde{A}_\mu^{a_L}
  +\sin^2{\theta_H\over 2}\,
    \partial_z \tilde{A}_\mu^{a_R}
   -{1\over \sqrt{2}} \sin\theta_H \,\partial_z 
  \myred{\tilde{A}_\mu^{\hat{a}}} ~,
\cr
  0&\!\!\!=\!\!\!&
   s_\phi\left(\sin^2{\theta_H\over 2}\,
  \partial_z \tilde{A}_\mu^{3_L}
  +\cos^2{\theta_H\over 2} \,\partial_z
   \tilde{A}_\mu^{3_R}
  +{1\over \sqrt{2}} \sin\theta_H \,\partial_z 
  \tilde{A}_\mu^{\hat{3}}\right)
  +c_\phi \partial_z B_\mu ~,
\cr
  0&\!\!\!=\!\!\!&
  \sin^2{\theta_H\over 2} \,\tilde{A}_\mu^{1_L,2_L}
  +\cos^2{\theta_H\over 2}\,\tilde{A}_\mu^{1_R,2_R}
  +{1\over \sqrt{2}} \sin\theta_H \,
\tilde{A}_\mu^{\hat{1},\hat{2}} ~,
\cr
  0&\!\!\!=\!\!\!&
  c_\phi
  \left(\sin^2{\theta_H\over 2} \,\tilde{A}_\mu^{3_L}
  +\cos^2{\theta_H\over 2} \,\tilde{A}_\mu^{3_R}
  +{1\over \sqrt{2}} \sin\theta_H \,
\tilde{A}_\mu^{\hat{3}}\right)
  -s_\phi B_\mu ~ .
\eea  

For the $W_\mu$ boson, the boundary conditions are
\bea
   0&\!\!\!=\!\!\!&
      -\sin\theta_H \,h_W^L
      +\sin\theta_H \,h_W^R
      -\sqrt{2}\cos\theta_H \,h_W^\wedge ,
      \label{bcw1}
\\
  0&\!\!\!=\!\!\!&
    \cos^2{\theta_H\over 2}\,
     \partial_z h_W^L
  +\sin^2{\theta_H\over 2}\,
    \partial_z h_W^R
   -{1\over \sqrt{2}} \sin\theta_H \,\partial_z 
  h_W^\wedge ,
    \label{bcw2}
\\
  0&\!\!\!=\!\!\!&
  \sin^2{\theta_H\over 2} \,h_W^L
  +\cos^2{\theta_H\over 2}\,h_W^R
  +{1\over \sqrt{2}} \sin\theta_H \,
h_W^\wedge .
   \label{bcw3}
\eea  
The \myred{conditions} (\ref{bcw1}) and (\ref{bcw3})
are automatically fulfilled with the wave functions 
(\ref{hw3}).
The  \myred{condition} (\ref{bcw2}) leads to Eq.~(\ref{bcw}).

For $A_\mu^\gamma$ boson,
the boundary conditions are
\bea
  0&\!\!\!=\!\!\!&
   s_\phi
   \partial_z h_\gamma
  +c_\phi \partial_z h_\gamma^B ,
  \label{agbc1}
\\
  0&\!\!\!=\!\!\!&
  c_\phi
  h_\gamma
  -s_\phi h_\gamma^B .
  \label{agbc2}
\eea
(\ref{agbc2}) is automatically fulfilled with (\ref{hg}). 
(\ref{agbc1}) leads to (\ref{agm}).

For $Z$ boson, \myred{the boundary conditions are}
\bea
  0&\!\!\!=\!\!\!&
   s_\phi\left(\sin^2{\theta_H\over 2}\,
  \partial_z h_Z^L
  +\cos^2{\theta_H\over 2} \,\partial_z
   h_Z^R
  +{1\over \sqrt{2}} \sin\theta_H \,\partial_z 
  h_Z^\wedge \right)
  +c_\phi \partial_z h_Z^B,
  \label{hzbc1}
\\
  0&\!\!\!=\!\!\!&
  c_\phi
  \left(\sin^2{\theta_H\over 2} \,h_Z^L
  +\cos^2{\theta_H\over 2} \,h_Z^R
  +{1\over \sqrt{2}} \sin\theta_H \,
h_Z^\wedge\right)
  -s_\phi h_Z^B.
    \label{hzbc2}
\eea
(\ref{hzbc2}) is automatically fulfilled with (\ref{hz}).
(\ref{hzbc1}) leads to (\ref{zm}).

\section{$SU(2)_L$ decomposition of  fermion currents}
\label{ap:fermi}

A fermion $\Psi$ in the vector representation of $SO(5)$
includes two $SU(2)_L$-doublets $\varphi_1, \varphi_2$
\beeq
\varphi_1 = \begin{pmatrix} \hat \psi_{11} \cr \hat \psi_{21}  \end{pmatrix}
~~,~~
\varphi_2 = \begin{pmatrix} \hat \psi_{12} \cr \hat \psi_{22}  \end{pmatrix}
\eneq
where $\hat \psi_{jk}$ is defined in (\ref{fermion1}). 
$(\varphi_1, \varphi_2) = (Q_1, q), (Q_2, Q_3)$ in (\ref{fermion2}).  
Bilinear operators for $\Psi$  are given in terms of $\varphi_1$, $\varphi_2$
as follows;
\bea
&&\hskip -1cm
\bar{\Psi}\gamma^\mu\, T^{a_L} \Psi =
       \bar{\varphi}_1 
     \gamma^\mu \tau_a \varphi_1 
     +\bar{\varphi}_2 
     \gamma^\mu \tau_a \varphi_2  ~,  \cr
\noalign{\kern 5pt}
&&\hskip -1cm
\bar{\Psi}\gamma^\mu\, T^{1_R} \Psi =
      {1\over 2} (\bar{\varphi}_1 
     \gamma^\mu \varphi_2 
     +\bar{\varphi}_2 
     \gamma^\mu \varphi_1)  ~,  \cr
\noalign{\kern 5pt}
&&\hskip -1cm
\bar{\Psi}\gamma^\mu\, T^{2_R} \Psi =
    -{i\over 2} (\bar{\varphi}_1 
     \gamma^\mu \varphi_2 
     -\bar{\varphi}_2 
     \gamma^\mu \varphi_1) ~, \cr
\noalign{\kern 5pt}
&&\hskip -1cm
  \bar{\Psi}\gamma^\mu\, T^{3_R} \Psi =
    {1\over 2} (\bar{\varphi}_1 
     \gamma^\mu \varphi_1 
     -\bar{\varphi}_2 
     \gamma^\mu \varphi_2) ~, \cr
\noalign{\kern 5pt}
&&\hskip -1cm
 \bar{\Psi}\gamma^\mu\, T^{\hat{1}} \Psi=
    -{1\over 2} ((\bar{\hat{\psi}}_{11}
    -\bar{\hat{\psi}}_{22}) \gamma^\mu\psi_5 
   +\bar{\psi}_5 \gamma^\mu(\hat{\psi}_{11}
      -\hat{\psi}_{22})
      )  ~, \cr
\noalign{\kern 5pt}
&&\hskip -1cm
\bar{\Psi}\gamma^\mu\, T^{\hat{2}} \Psi =
     {i\over 2} ((\bar{\hat{\psi}}_{11}
    +\bar{\hat{\psi}}_{22}) \gamma^\mu\psi_5 
   -\bar{\psi}_5 \gamma^\mu(\hat{\psi}_{11}
      +\hat{\psi}_{22})  ) ~, \cr
\noalign{\kern 5pt}
&&\hskip -1cm
\bar{\Psi}\gamma^\mu\, T^{\hat{3}} \Psi =
     {1\over 2} ((\bar{\hat{\psi}}_{12}
    +\bar{\hat{\psi}}_{21}) \gamma^\mu\psi_5 
   +\bar{\psi}_5 \gamma^\mu(\hat{\psi}_{12}
      +\hat{\psi}_{21})  ) ~, \cr
\noalign{\kern 5pt}
&&\hskip -1cm
 \bar{\Psi}\gamma^\mu\, T^{\hat{4}} \Psi =
    - {i\over 2} ((\bar{\hat{\psi}}_{12}
    -\bar{\hat{\psi}}_{21}) \gamma^\mu\psi_5 
   -\bar{\psi}_5 \gamma^\mu(\hat{\psi}_{12}
      -\hat{\psi}_{21})  ) ~.
\label{decomposition1}
\eeqn
Here $\tau_a ={\sigma_a/ 2}$, $a=1,2,3$.  
Other useful formulas are 
\beqn
&&\hskip -1cm
\bar{\Psi}\gamma^\mu\, Q_E \Psi =
    \bar{\varphi}_1 \gamma^\mu
      (\tau_3  +{1\over 2} +Q_X)  \varphi_1
    + \bar{\varphi}_2 \gamma^\mu
      (\tau_3   -{1\over 2} +Q_X)  \varphi_2
   +  \bar{\psi}_5 \gamma^\mu   Q_X   \psi_5 ~,  \cr
\noalign{\kern 5pt}
&&\hskip -1cm
\bar{\Psi}\gamma^\mu\, T^{\mp_L} \Psi =
    \bar{\varphi}_1   \gamma^\mu   \tau^\mp  \varphi_1 
 +\bar{\varphi}_2   \gamma^\mu   \tau^\mp  \varphi_2  ~, \cr
\noalign{\kern 5pt}
&&\hskip -1cm
\bar{\Psi}\gamma^\mu\, T^{-_R} \Psi =
    {1\over \sqrt{2}}    \bar{\varphi}_2 \gamma^\mu\varphi_1 ~ , \cr
\noalign{\kern 5pt}
&&\hskip -1cm
   \bar{\Psi}\gamma^\mu\, T^{+_R} \Psi =
   {1\over \sqrt{2}}   \bar{\varphi}_1 \gamma^\mu\varphi_2  ~ , \cr
\noalign{\kern 5pt}
&&\hskip -1cm
\bar{\Psi}\gamma^\mu\, T^{\hat{-}} \Psi =
    {1\over \sqrt{2}}     (\bar{\hat{\psi}}_{22} \gamma^\mu\psi_5 
      -\bar{\psi}_5\gamma^\mu\hat{\psi}_{11}     ) ~ ,    \cr
\noalign{\kern 5pt}
&&\hskip -1cm
\bar{\Psi}\gamma^\mu\, T^{\hat{+}} \Psi =
    {1\over \sqrt{2}}    (-\bar{\hat{\psi}}_{11} \gamma^\mu\psi_5 
      +\bar{\psi}_5\gamma^\mu\hat{\psi}_{22}  ) ~ .      
\label{decomposition2}
\eeqn
Here  $\tau^{\mp} ={(\tau_1\mp i\tau_2)/ \sqrt{2}}$.

Thus the electroweak couplings for $\Psi$ in the vector representation are
given by
\beqn
&&\hskip -1cm
\bar{\Psi} \gamma^\mu  (g_A \tilde{A}_\mu + g_B Q_X \tilde{B}_\mu)  \Psi \cr
\noalign{\kern 5pt}
&&\hskip -1cm
=  e A_\mu^\gamma (x) J_\gamma^\mu
   + g_A W_\mu (x) J_W^\mu + g_A W_\mu^\dagger  (x) J_{W^\dagger }^\mu
   + \frac{g_A}{\cos \theta_W} Z_\mu (x) J_Z^\mu
\label{EWcoupling1}
\eeqn
where  
\beqn
&&\hskip -1cm
J_\gamma^\mu = 
\bar{\varphi}_1 \gamma^\mu({\tau_3}
     +{1\over 2} +Q_X )\varphi_1
   +\bar{\varphi}_2 \gamma^\mu   ({\tau_3} -{1\over 2}   +Q_X)\varphi_2
   +\bar{\psi}_5 \gamma^\mu Q_X\psi_5   ~, \cr
\noalign{\kern 10pt}
&&\hskip -1cm
J_W^\mu = N_W \bigg\{   
     \bar{\varphi}_1 \gamma^\mu \tau^-   \varphi_1
     +\bar{\varphi}_2 \gamma^\mu \tau^-     \varphi_2 
- \frac{1- \cos\theta_H}{2} \Big[  \bar{\varphi}_1 \gamma^\mu \tau^- \varphi_1
   +\bar{\varphi}_2 \gamma^\mu \tau^- \varphi_2
   -{1\over \sqrt{2}}\bar{\varphi}_2 \gamma^\mu \varphi_1 \Big] \bigg\} 
\cr
\noalign{\kern 10pt}
&&\hskip 1cm
- D_W \frac{\sin\theta_H}{2} 
    \Big\{   \bar{\hat{\psi}}_{22}   \gamma^\mu  \psi_5
     -\bar{\psi}_5 \gamma^\mu  \hat{\psi}_{11} \Big\}  ~, 
\cr
\noalign{\kern 10pt}
&&\hskip -1cm
J_{W^\dagger }^\mu = N_W \bigg\{
    \bar{\varphi}_1 \gamma^\mu \tau^+  \varphi_1
     +\bar{\varphi}_2 \gamma^\mu \tau^+   \varphi_2
-\frac{1- \cos\theta_H}{2}\Big[  \bar{\varphi}_1 \gamma^\mu \tau^+ \varphi_1
   +\bar{\varphi}_2 \gamma^\mu \tau^+ \varphi_2
   -{1\over \sqrt{2}}\bar{\varphi}_1 \gamma^\mu \varphi_2 \Big] \bigg\}
\cr
\noalign{\kern 10pt}
&&\hskip 1cm
- D_W \frac{\sin\theta_H}{2} 
 \Big\{  -\bar{\hat{\psi}}_{11} \gamma^\mu \psi_5
     +\bar{\psi}_5 \gamma^\mu  \hat{\psi}_{22} \Big\} ~, 
\cr
\noalign{\kern 10pt}
&&\hskip -1cm
J_Z^\mu = N_Z \bigg\{
 \bar{\varphi}_1 \gamma^\mu{\tau_3} \varphi_1
     +\bar{\varphi}_2 \gamma^\mu{\tau_3}\varphi_2
 - \frac{1- \cos\theta_H}{2} \Big[ \bar{\varphi}_1 \gamma^\mu ({\tau_3}
     -{1\over 2}) \varphi_1
     +\bar{\varphi}_2 \gamma^\mu ({\tau_3}  +{1\over 2}) \varphi_2 \Big] \bigg\}
\cr
\noalign{\kern 10pt}
&&\hskip 1cm
- D_Z \frac{\sin\theta_H}{2} 
\Big\{   (\bar{\hat{\psi}}_{12}  +\bar{\hat{\psi}}_{21})\gamma^\mu  \psi_5
     +\bar{\psi}_5 \gamma^\mu  (\hat{\psi}_{12}+\hat{\psi}_{21})  \Big\}
     - N_Z \sin^2 \theta_W J_\gamma^\mu ~.
\label{EWcoupling2}
\eeqn

\section{Electroweak currents}
 \label{ap:gauge-couple}

In evaluating the electroweak currents in Section 5, the following formulas
are useful.
The tilde over fields for denoting the twisted gauge is omitted in this appendix.

\subsubsection*{(i) Photon couplings}

4D Lagrangian  for \myred{the} photon-top coupling \myred{is} given by
\bea
  \int_1^{z_L}
   {dz \over k}
  \, {2\over 3}eA_\mu^\gamma
  \left[ \bar{U} \gamma^\mu\, U 
    +\overline{B+t\over \sqrt{2}} \gamma^\mu
        {B+t\over \sqrt{2}}
          +\overline{B-t\over \sqrt{2}}\gamma^\mu
          {B-t\over \sqrt{2}}
     +\overline{t'} \gamma^\mu t'
    \right] .
\eea
For \myred{the} photon-bottom coupling one finds
\bea
 \int_1^{z_L}
   {dz \over k}
  \, \left(-{1\over 3}\right) eA_\mu^\gamma
  \left[ 
   \bar{b}\gamma^\mu\,
     b
  +\overline{D+X\over \sqrt{2}}\gamma^\mu
      {D+X\over \sqrt{2}}
 +\overline{D-X\over \sqrt{2}}\gamma^\mu
   {D-X\over \sqrt{2}}
 +\overline{b'}\gamma^\mu b' 
  \right] .
\eea
For \myred{the} photon-electron coupling one finds
\bea
 \int_1^{z_L}
   {dz \over k}
  \, \left(-{1}\right) eA_\mu^\gamma
  \left[ 
   \overline{E+L_X\over \sqrt{2}}\gamma^\mu
      {E+L_X\over \sqrt{2}}
 +\overline{E-L_X\over \sqrt{2}}\gamma^\mu
   {E-L_X\over \sqrt{2}}
 +\overline{e'}\gamma^\mu e' 
  \right] .
\eea
These lead to Eq.\ (\ref{photoncoupling1}).  

\subsubsection*{(ii) $W$ boson couplings}

Lagrangian terms for \myred{the} $Wt\bar{b}$ coupling are
\bea
&&\hskip -1cm
   \int_1^{z_L} {dz \over k} 
   \, {1\over 2} g_A W_\mu\left[
   N_W
  \left\{ \overline{D+X\over \sqrt{2}}
  \gamma^\mu U
  +\overline{D-X\over \sqrt{2}}\gamma^\mu U
   +\bar{b} \gamma^\mu  {B+t\over \sqrt{2}}
   -\bar{b} \gamma^\mu {B-t\over \sqrt{2}}\right\}
    \right. \cr
\noalign{\kern 10pt}
&&\hskip -.8cm
 \left.
  - (1-\cos\theta_H)
       N_W \left\{
          \overline{D-X\over \sqrt{2}}\gamma^\mu U
      -\bar{b} \gamma^\mu 
      {B-t\over \sqrt{2}}\right\}
   - \sin\theta_H
     D_W \left\{ \bar{b}\gamma^\mu\, t'
      -\overline{b'} \gamma^\mu\,U\right\}
  \right] .
\eea
This leads to Eq.\ (\ref{Wcoupling2}).

\subsubsection*{(iii) $Z$ boson couplings}

Lagrangian terms for \myred{the} $Zt\bar{t}$ coupling are
\bea
&&\hskip -1cm
\int_1^{z_L} {dz\over k} \,
 {g_A Z_\mu\over \cos\theta_W}
      \left[ {N_Z \over 2}\left(
       \bar{U}\gamma^\mu  U
           - \overline{B+t\over\sqrt{2}}\gamma^\mu
         {B-t\over \sqrt{2}}
      - \overline{B-t\over \sqrt{2}}\gamma^\mu
        {B+t\over \sqrt{2}}
   \right)
   \right.  \cr
\noalign{\kern 10pt}
&&\hskip -.5cm
    -{2\over 3} \sin^2\theta_W N_Z \bigg\{
  \bar{U} \gamma^\mu U 
    +\overline{B+t\over \sqrt{2}} \gamma^\mu
        {B+t\over \sqrt{2}}
          +\overline{B-t\over \sqrt{2}}\gamma^\mu
           {B-t\over \sqrt{2}}
     +\overline{t'} \gamma^\mu t'
    \bigg\}  \cr
\noalign{\kern 10pt}
&&\hskip -.5cm
+ {(1-\cos \theta_H)\over 2}
       N_Z \left\{
         \overline{B+t\over \sqrt{2}}\gamma^\mu
            {B-t\over \sqrt{2}}
      +\overline{B-t\over \sqrt{2}}\gamma^\mu
          {B+t\over \sqrt{2}}
           \right\}  \cr
\noalign{\kern 10pt}
&&\hskip -.5cm
 \left.
     -{\sin\theta_H\over 2} D_Z
  \left\{
    \overline{B+t\over \sqrt{2}}\gamma^\mu\,
      t'
      +\overline{t'} \gamma^\mu\,{B+t\over\sqrt{2}}
    \right\}
 \right] .
  \label{Zc1}
\eea
Lagrangian terms for \myred{the} $Zb\bar{b}$ coupling are
\bea
&&\hskip -1cm
 \int_1^{z_L} {dz\over kz}
 \, {g_A Z_\mu \over \cos\theta_W}
     \left[ -{N_Z\over 2}
     \left(  \bar{b}\gamma^\mu b
   +\overline{D+X\over \sqrt{2}}\gamma^\mu
      {D-X\over \sqrt{2}}
      +\overline{D-X\over \sqrt{2}}\gamma^\mu
        {D+X\over \sqrt{2}}
      \right)  \right.  \cr
\noalign{\kern 10pt}
&&\hskip -.5cm
    +{1\over 3}\sin^2\theta_W N_Z \bigg\{
   \bar{b}\gamma^\mu b
  +\overline{D+X\over \sqrt{2}}\gamma^\mu
      {D+X\over \sqrt{2}}
 +\overline{D-X\over \sqrt{2}}\gamma^\mu
   {D-X\over \sqrt{2}}
 +\overline{b'}\gamma^\mu b'\bigg\}  \cr
\noalign{\kern 10pt}
&&\hskip -.5cm
+{(1-\cos \theta_H)\over 2} N_Z
       \left\{
    \overline{D+X\over \sqrt{2}}\gamma^\mu
      {D-X\over \sqrt{2}}
       +\overline{D-X\over \sqrt{2}}\gamma^\mu
          {D+X\over \sqrt{2}}
           \right\}  \cr
\noalign{\kern 10pt}
&&\hskip -.5cm
 \left.
     -{\sin\theta_H\over 2} D_Z
  \left\{
    \overline{D+X\over \sqrt{2}}\gamma^\mu\,
       b'
    +\overline{b'} \gamma^\mu\,{D+X\over \sqrt{2}}\right\}
 \right] .
 \label{Zc2}
\eea
Eq.\   (\ref{Zcoupling2}) follows from  (\ref {Zc1}) and  (\ref {Zc2}).

\section{Gauge couplings determined with
inputs of the  running masses of quarks and leptons
at the $m_Z$ scale}
\label{app:running}

In this appendix, we summarize the results for the coefficients of mode functions
and the gauge couplings  
with input parameters corresponding to
the running masses of quarks and leptons at the $m_Z$ scale 
given \myred{in  ref.~\cite{XZZ}.}
Here we treat the case of  massless neutrinos.

%
%

\begin{table}[htb]
\begin{center}
\caption{The quark and lepton masses at the $m_Z$ scale are input parameters, 
from which
$c$, $|\tilde \mu/\mu_2|$, 
and the coefficients of the wave functions are determined.
Here  $z_L=10^{15}$, $k=4.7\times 10^{17}$, 
$\theta_H ={\pi\over 2}$ and $\theta_W=0.2312$.
}
\label{tab:cmu-x}
\vskip 10pt
\begin{tabular}{|c|cccc|}
\hline 
& $c$ & $|\tilde{\mu}/\mu_2|$ & 
 \multicolumn{2}{c|}{Mass (GeV)} \\ \hline
 $W$,$Z$ & & & 80.398 & 91.1876  \\ \hline
$u$,$d$ & 0.8436 & 4.000
   &  $1.27 \times 10^{-3}$ & $2.90 \times 10^{-3}$ \\
$c$,$s$ & 0.6795 & 0.4679
        & 0.619 & 0.055 \\
$t$,$b$ & 0.4325 & 0.3249
        & 171.7 & 2.89 \\ 
\hline 
$\nu_e$,$e$ & 0.9007 & --- 
        &  0. & $0.486570161\times 10^{-3}$ \\
$\nu_\mu$,$\mu$ & 0.7362 & ---
        &  0. & 0.1027181359 \\
$\nu_\tau$,$\tau$ & 0.6462 & ---
         & 0. & 1.74624 \\ 
   \hline
\end{tabular}
\vskip 10pt
\begin{tabular}{|c|c|c|cc|}
\hline 
& \multicolumn{2}{c}{$a_U$} & $a_{B+t}$ & $a_{t'}$ \\ \hline
$u$ & 
\multicolumn{2}{c}{
$-1.683\times 10^{-13}$} & 
$5.213\times 10^{-14}$  & 
$5.184\times 10^{-8}$     
\\ %
$c$ &
\multicolumn{2}{c}{
$-3.407\times 10^{-12}$} & 
$2.712\times 10^{-11}$  & 
$4.857\times 10^{-8}$     
\\ %
$t$ &
\multicolumn{2}{c}{
$-2.136\times 10^{-10}$} & 
$8.757\times 10^{-9}$   & 
$4.387\times 10^{-8}$     
\\
\hline
& \multicolumn{2}{c}{$a_b$} & $a_{D+X}$ & $a_{b'}$ \\ \hline
$d$ &
\multicolumn{2}{c}{
$-7.372\times 10^{-14}$} & 
$1.190\times 10^{-13}$  & 
$-5.184\times 10^{-8}$    
\\ %
$s$ &
\multicolumn{2}{c}{
$-3.835\times 10^{-11}$} & 
$2.409\times 10^{-12}$  & 
$-4.857\times 10^{-8}$    
\\ %
$b$ &
\multicolumn{2}{c}{
$-1.168\times 10^{-8}$}  & 
$1.424\times 10^{-10}$  & 
$-4.319\times 10^{-8}$    
\\ \hline
\noalign{\kern 10pt}
\hline 
& $a_{\nu_\tau}$ & 
& $a_{\tau + L_{1X}}$ & $a_{\tau'}$ \\ \hline
$\nu_e$ & 
$2.766\times 10^{-14}$  & 
 $e$ & 
$1.956\times 10^{-14}$  & 
$-5.293\times 10^{-8}$    
\\ %
$\nu_{\mu}$ &
$6.216\times 10^{-12}$  & 
$\mu$ 
& 
$4.395\times 10^{-12}$  & 
$-4.972\times 10^{-8}$    
\\ %
$\nu_{\tau}$ &
$1.097\times 10^{-10}$ & 
$\tau$ &
$7.760\times 10^{-11}$  & 
$-4.788\times 10^{-8}$    
 \\
\hline
\end{tabular}
\end{center}
\end{table}
%

%
%
%
%

\newpage 

For the input parameters given in Table~\ref{tab:cmu-x},
the normalized coefficients are obtained as
in Table~\ref{tab:aprime-x}.

\begin{table}[htb]
\begin{center}
\caption{Normalized coefficients in mode expansion 
for quarks and leptons.}
\label{tab:aprime-x}
\vskip 10pt
\begin{tabular}{|c|ccc|ccc|}
\hline 
& $a'_{UL}$ & $a'_{B+tL}$ & $a'_{t'L}$ 
& $a'_{UR}$ & $a'_{B+tR}$ & $a'_{t'R}$
\\ \hline
$u$ & 
$-0.9160$               & 
0.2837                  & 
0.2837                  & 
$-7.658\times 10^{-12}$ & 
$2.371\times 10^{-12}$  & 
1  
\\ 
$c$ &
$-0.08850$             & 
0.7043                 & 
0.7043                 & 
$-7.213\times 10^{-8}$ & 
$5.740\times 10^{-7}$  & 
1
\\
$t$ &
$-0.01773$             & 
0.7267                 & 
0.6867                 & 
$-0.001221$            & 
0.05007                & 
0.9987                   
\\
\hline
\noalign{\kern 5pt}
\hline
& $a'_{bL}$ & $a'_{D+XL}$ & $a'_{b'L}$
& $a'_{bR}$ & $a'_{D+XR}$ & $a'_{b'R}$ 
\\ \hline
$d$ & 
$-0.4012$               & 
0.6477                  & 
$-0.6477$               & 
$-7.658\times 10^{-12}$ & 
$1.237\times 10^{-11}$  & 
$-1$
\\ 
$s$ &
$-0.9961$              & 
0.06258                &  
$-0.06258$             & 
$-7.213\times 10^{-8}$ & 
$4.532\times 10^{-9}$  & 
$-1$
\\ 
$b$ &
$-0.9999$   & 
0.01219     & 
$-0.01156$  & 
$-0.001138$ & 
0.00001388  & 
$-1$            
\\
\hline
\end{tabular}
%
\vskip 10pt
\begin{tabular}{|c|c|}
\hline  
& $a'_{\nu_\tau L}$ 
\\ \hline
$\nu_e$ & 
1 
\\ 
$\nu_{\mu}$ &
1 
\\ 
$\nu_{\tau}$ &
1
\\
\hline
\end{tabular}
\hspace{-0.5ex}
\begin{tabular}{|c|cccc|}
\hline
& $a'_{\tau + L_{1X} L}$ & $a'_{\tau' L}$
& $a'_{\tau +L_{1X} R}$ & $a'_{\tau' R}$ 
\\ \hline 
$e$ &
0.7071                      & 
$-0.7071$                   & 
$3.494\times\!\! 10^{-13}$  & 
$-1$
\\ 
$\mu$ & 
0.7071                      & 
$-0.7071$                   & 
$1.563\times\!\! 10^{-8}$   & 
$-1$
\\ 
$\tau$ & 
0.7071                      & 
$-0.7071$                   & 
$4.608\times\!\! 10^{-6}$   & 
$-1$
\\
\hline
\end{tabular}
\end{center}
\end{table}
%

%
%

\newpage 

For the input parameters given in Table~\ref{tab:cmu-x},
the $W$ couplings and the normalized $W$ couplings are given
in Table~\ref{tab:gw-x} and Table~\ref{tab:gw2-x}, respectively.

\begin{table}[htb]
\begin{center}
\caption{The $W$ couplings $g^{(W)}_{f, LR}$ and $g'^{(W)}_{f, LR}$ 
at $\theta_H=\onehalf \pi$.  
}\label{tab:gw-x}
\vskip 10pt
\begin{tabular}{|c|ccc|}
\hline 
      $f$      & $u,d$   & $c,s$  & $t,b$ \\ \hline
$g^{(W)}_{fL}/\myred{\bar{g}_A}$ 
& 1.00533 & 1.00533 & 0.9818 \\ 
$g^{(W)}_{fR}/\myred{\bar{g}_A}$ &
$-4.458\times\!\! 10^{-12}$ & 
$-4.800\times\!\! 10^{-8}$  & 
$-0.0008771$   
 \\ 
  \hline
  $g'^{(W)}_{fL}$ & 1. & 1. & 0.9766 \\ 
$g'^{(W)}_{fR}$ & 
$-4.434\times\!\! 10^{-12}$ & 
$-4.774\times\!\! 10^{-8}$  & 
$-0.0008725$ 
\\ 
\hline
\noalign{\kern 7pt}
\hline
       $f$     & $\nu_e,e$   & $\nu_{\mu},\mu$  & $\nu_{\tau},\tau$ \\ \hline
$g^{(W)}_{fL}/\myred{\bar{g}_A}$ 
& 1.00533 & 1.00533 & 1.00533
 \\ 
  \hline
  $g'^{(W)}_{fL}$ & 1. & 1. & 1.
\\ 
\hline
\end{tabular}
\end{center}
\end{table}
%

%
%

%
\begin{table}[hbt]
\begin{center}
\caption{$g'^{(W)}-1$.
}
\label{tab:gw2-x}
\vskip 1pt
\begin{tabular}{|c|ccccc|}
\hline 
 $f$   & $\nu_{\mu},\mu$  & $\nu_{\tau},\tau$ & $u,d$ & $c,s$  & $t,b$  \\ \hline
&
$-9.669\times\!\! 10^{-9}$  & 
$-2.643\times\!\! 10^{-6}$   & 
$-9.871\times\!\! 10^{-12}$ & 
$-3.409\times\!\! 10^{-7}$  & 
$-0.02341$    
 \\ \hline
\end{tabular}
\end{center}
\end{table}
%

%
%

\newpage

For the input parameters given in Table~\ref{tab:cmu-x},
the $Z$ couplings of quarks are obtained as
in Table~\ref{tab:gz-x}.

\begin{table}[htb]
\begin{center}
\caption{$Z$ boson couplings of quarks at $\theta_H = \onehalf \pi$.
} 
\label{tab:gz-x}
\vskip 10pt
\begin{tabular}{|c|ccc|c|} 
\hline 
 $f$ & $u$ & $c$ & $t$  & SM \\ \hline
$g^{(Z)}_{fL}/\myred{\bar{g}_A}$  & 0.3483 & 
0.3483 & 
0.3483 & 
0.3220   
 \\
$g^{(Z)}_{fR}/\myred{\bar{g}_A}$ & $-0.1564$ & 
$-0.1564$ & 
$-0.1564$ & 
$-0.1835$   
 \\
$g^{(Z)3}_{fL}/\myred{\bar{g}_A}$ &
1.00699 & 
1.00699 & 
0.9546    
 & 1
 \\
$g^{(Z)3}_{fR}/\myred{\bar{g}_A}$ &
$-1.957\times\!\! 10^{-12}$ & 
$-5.414\times\!\! 10^{-7}$  & 
$-0.05434$                    
 & 0
\\
$g^{(Z)Q}_{fL}/\myred{\bar{g}_A}$  &
1.00699 & 
1.00699 & 
1.0079     
 & 1
\\
$g^{(Z)Q}_{fR}/\myred{\bar{g}_A}$ &
1.0148 & 
1.0145 & 
1.0141   
 &1
 \\
 \hline
 \noalign{\kern 5pt}
 \hline
 $f$ &  $d$ & $s$ & $b$  & SM \\ \hline
$g^{(Z)}_{fL}/\myred{\bar{g}_A}$  
  & $-0.4259$ & 
$-0.4259$ & 
$-0.4264$   
 & $-0.4229$ \\
$g^{(Z)}_{fR}/\myred{\bar{g}_A}$ 
 & 0.07820 & 
0.07819&  
0.07815   
  & 0.07707 \\
$g^{(Z)3}_{fL}/\myred{\bar{g}_A}$
 & 1.00699 & 
1.00699 & 
1.00815   
 & 1\\
$g^{(Z)3}_{fR}/\myred{\bar{g}_A}$
 & $-1.020\times\!\! 10^{-11}$ & 
$-4.274\times\!\! 10^{-9}$  & 
$-0.00001422$                 
 & 0
\\
$g^{(Z)Q}_{fL}/\myred{\bar{g}_A}$ 
  & 1.00699 & 
1.00699 & 
1.00816   
 & 1
\\
$g^{(Z)Q}_{fR}/\myred{\bar{g}_A}$
  &1.0148 & 
1.0145 & 
1.0140   
 & 1 \\
 \hline 
\end{tabular}
\end{center}
\end{table}
%

%
%

\newpage 

For the input parameters given in Table~\ref{tab:cmu-x},
the $Z$ couplings of leptons are obtained as
in Table~\ref{tab:gz-l-x}.

\begin{table}[htb]
\begin{center}
\caption{$Z$ boson couplings of leptons at $\theta_H = \onehalf \pi$.
} 
\label{tab:gz-l-x}
\vskip 10pt
\begin{tabular}{|c|ccc|c|} 
\hline 
 $f$ & $\nu_e$ & $\nu_\mu$ & $\nu_\tau$  & SM \\ \hline
$g^{(Z)}_{fL}/\myred{\bar{g}_A}$  & 
0.5035 & 
0.5035 & 
0.5035   
 & 0.5 \\
$g^{(Z)3}_{fL}/\myred{\bar{g}_A}$ & 
1.00699 & 
1.00699 & 
1.00699   
 & 1
 \\
$g^{(Z)Q}_{fL}/\myred{\bar{g}_A}$  &
1.00699 & 
1.00699 & 
1.00699   
 & 1
\\
 \hline
 \noalign{\kern 5pt}
 \hline
 $f$ &  $e$ & $\mu$ & $\tau$  & SM \\ \hline
$g^{(Z)}_{fL}/\myred{\bar{g}_A}$  & 
$-0.2707$ & 
$-0.2707$ & 
$-0.2707$   
 & $-0.2688$ \\
$g^{(Z)}_{fR}/\myred{\bar{g}_A}$  & 
$0.2346$ & 
$0.2346$ & 
$0.2345$   
  & 0.2312 \\
$g^{(Z)3}_{fL}/\myred{\bar{g}_A}$ & 
1.00699 & 
1.00699 & 
1.00698   
 & 1\\
$g^{(Z)3}_{fR}/\myred{\bar{g}_A}$ & 
$-2.736\times\!\! 10^{-13}$ & 
$-1.411\times\!\! 10^{-8}$  & 
$-4.455\times\!\! 10^{-6}$    
 & 0
\\
$g^{(Z)Q}_{fL}/\myred{\bar{g}_A}$   & 
1.00699 & 
1.00699 & 
1.00699   
 & 1
\\
$g^{(Z)Q}_{fR}/\myred{\bar{g}_A}$  & 
1.0149 & 
1.0146 & 
1.0145   
 & 1 \\
 \hline 
\end{tabular}
\end{center}
\end{table}
%

%
%

\newpage 

For the input parameters given in Table~\ref{tab:cmu-x},
the normalized $Z$ couplings are obtained as
in Table~\ref{tab:gz2-x}.

\begin{table}[htb]
\begin{center}
\caption{$Z$ boson couplings $g'^{(Z)}_{fL,R}$ in (\ref{Zcoupling3}) 
and the deviation from SM.  The quantity (\ref{Zcoupling4}), which measures the 
deviation from the standard model,  is given in parentheses.}
\label{tab:gz2-x}
\vskip 10pt
\begin{tabular}{|c|ccc|c|} 
\hline 
 $f$ & $u$ & $c$ & $t$ & SM  \\ \hline
$g'^{(Z)}_{fL}$  & 
0.3464 & 
0.3464 & 
0.3202 & 
0.3459   
\\
& 
(+0.001644) & 
(+0.001643) & 
$(-0.07408)$ & 
 \\
$g'^{(Z)}_{fR}$
 & $-0.1556$ & 
$-0.1555$ & 
$-0.1825$ & 
$-0.1541$   
\\
& 
(+0.009406) & 
(+0.009137) & 
(+0.1840)   & 
\\ \hline
\noalign{\kern 5pt}
\hline
$f$  & $d$ & $s$ & $b$ & SM  \\ \hline
$g'^{(Z)}_{fL}$  & 
$-0.4236$ & 
$-0.4236$ & 
$-0.4241$ & 
$-0.4229$   
\\
& 
(+0.001644) & 
(+0.001644) & 
(+0.002799) & 
 \\
$g'^{(Z)}_{fR}$ & 
0.07779 & 
0.07777 & 
0.07774 & 
0.07707   
\\
& 
(+0.009406) & 
(+0.009135) & 
(+0.008683) & 
\\ \hline
\noalign{\kern 5pt}
\hline 
 $f$ & $\nu_e$ & $\nu_{\mu}$ & $\nu_{\tau}$ & SM  \\ \hline
$g'^{(Z)}_{fL}$  & 
0.500822 & 
0.500822 & 
0.500822 & 
0.5
\\
& 
(+0.001644) & 
(+0.001644) & 
(+0.001644) & 
\\ \hline
\noalign{\kern 5pt}
\hline 
 $f$ & $e$ & $\mu$ & $\tau$ & SM  \\ \hline
$g'^{(Z)}_{fL}$  & 
$-0.2692$ & 
$-0.2692$ & 
$-0.2692$ & 
$-0.2688$   
\\
& 
(+0.001644) & 
(+0.001644) & 
(+0.001633) & 
 \\
$g'^{(Z)}_{fR}$
 & 
0.2334 & 
0.2333 & 
0.2333 & 
0.2312   
\\
& 
(+0.009485) & 
(+0.009234) & 
(+0.009082) & 
\\ \hline
\end{tabular}
\end{center}
\end{table} 

\end{appendix}

\newpage


\renewenvironment{thebibliography}[1]
         {\begin{list}{[$\,$\arabic{enumi}$\,$]}  
         {\usecounter{enumi}\setlength{\parsep}{0pt}
          \setlength{\itemsep}{0pt}  \renewcommand{\baselinestretch}{1.2}
          \settowidth
         {\labelwidth}{#1 ~ ~}\sloppy}}{\end{list}}



\end{document}